\documentclass[10pt,final,a4paper,twocolumn,twoside,journal,romanappendices]{IEEEtran}
\usepackage{cite}
\usepackage{epstopdf}
\usepackage{epsfig}
\usepackage{amsthm}
\usepackage{amsmath,amssymb,amsfonts,amstext,amsbsy,amsopn,dsfont}
\usepackage{cases}
\usepackage{sublabel}
\usepackage{color}
\usepackage{array}
\usepackage{bigints}

\newtheorem{theorem}{\textbf{Theorem}}

\newtheorem{lemma}{\textbf{Lemma}}

\newtheorem{definition}{\textbf{Definition}}

\newcommand{\defn}{\triangleq}
\newcommand{\dif}{\textmd{d}}
\newcommand{\gc}{\mathsf{gc}}
\newcommand{\mr}{\mathsf{mr}}
\newcommand{\nb}{\mathsf{nb}}
\newcommand{\subU}{\mathsf{u}}

\newcommand{\on}{\mathsf{on}}
\newcommand{\off}{\mathsf{off}}
\newcommand{\subVoid}{\emptyset}

\newcommand{\argsup}{\arg\sup}

\begin{document}

\title{Fundamentals of the Downlink Green Coverage and Energy Efficiency in Heterogeneous Networks}


\author{Chun-Hung Liu, \IEEEmembership{Senior Member, IEEE} and Kok Leong Fong
\thanks{The work of C.-H. Liu was supported in part by the Ministry of Science and Technology of Taiwan with grant 104-2628-E-009-006-MY3. Part of this paper was presented in IEEE Globecom, December 2016 \cite{CHL16}.}%
\thanks{C.-H. Liu is with the Department of Electrical and Computer Engineering, National Chiao Tung University, Hsinchu Taiwan  (e-mail: chungliu@nctu.edu.tw) and K. L. Fong was in the international graduate program of Electrical Engineering and Computer Science (EECS), National Chiao Tung University from September 2014 to March 2016.}}

\maketitle

\begin{abstract}
This paper studies the proposed green (energy-efficient) coverage probability, link and network energy efficiencies in the \textit{downlink} of a heterogeneous cellular network (HetNet) consisting of $K$ independent Poisson point processes (PPPs) of base stations (BSs). The important statistical properties of the universal (general) cell association functions are first studied and the cell load statistics for power-law cell association functions, which can characterize the accurate void cell probability of a BS in every tier, is also derived.  A simple and feasible green channel-aware cell association (GCA) scheme is proposed and the green coverage probability is also proposed for any particular cell association scheme, such as the maximum received power association (MRPA) and nearest base station association (NBA) schemes. Then the link and network energy efficiencies are proposed to characterize the mean spectrum efficiency per unit power consumption for a BS and the mean area spectrum efficiency for a HetNet, respectively. All the tight bounds on the green coverage probability, link and network energy efficiencies for the GCA, MRPA and NBA schemes are found. They are theoretically shown to pose the fundamental maximum limits on the link and network energy efficiencies achieved by any other cell association schemes and such a fact is validated by numerical results as well.
\end{abstract}


\section{Introduction}
In recent years, the longstanding uniform topology of a cellular network has been gradually vanishing and transformed into a complex and heterogeneous network (HetNet) consisting of different types of base stations (BSs), such as macro, micro, small cell (e.g., picocell and femtocell) BSs, etc. Such a HetNet is potentially able to support a huge amount of traffic flow from the explosive proliferations of smart wireless handsets since its network capacity seems not to reach a hard limit even when more and more BSs are deployed in a HetNet as long as BSs are well positioned without causing severe interference between them \cite{AGNMO12}\cite{JGANDREWS13}. Deploying more small cell BSs instead of macro BSs in a HetNet not only brings considerable network capacity but also consumes less power since small cell BSs have much less power consumption than macro BSs. However, some practical problems would arise in the dense HetNet, such as, operation and maintenance expenditure, traffic management and offloading, cell load balancing, energy consumption and management, etc., and how to leverage these problems to further improve the capacity and low power consumption benefits is an important issue that needs to be exhaustively investigated. 

\subsection{Prior Work and Motivation}
HetNets with a dense small cell topology definitely help the entire information and communication technology (ICT) industry control the CO$_2$ emission, which is estimated to account for at least 6\% of the global emission in 2020 \cite{AFGFJMGB11,ZHHBVKB11}. Effective energy-saving methods are certainly worth exploring in a dense HetNet since power control and operation at small cell BSs are much easier than macro BSs \cite{IAFBLH11}. The majority of current works on energy-saving techniques for HetNets focus on the BS-level management such as sleep/active mode control, cell activation, scheduling, etc. (see some typical works in \cite{WWGS10,SRCWC13,YSSTQSQMKHS13,JPHTPHKX13,EMDKCS15,CTPCHLLCW15,CHLLCW1503,CHLLCW16}). The main idea in these related prior works is to save power by switching BSs to the sleep/dormant mode if they satisfy certain conditions, such as no users, carrying low traffic and meeting the predesignated constraints on power consumption, etc. 

From the green communications point of view, offloading traffic from macro BSs to small cell BSs should improve the (bits-per-joule) energy efficiency of BSs while serving users. However, how to do traffic allocation/offloading between different types of BSs to accomplish a certain level of energy efficiency is hardly investigated in the literatures. Traffic flow behaviors in a HetNet are significantly affected by the cell (user) association strategies. If there exists an energy-efficient cell association means to make all the traffic flow into BSs with high energy efficiency, saving huge amount of energy can be certainly anticipated. Most prior works on cell association and traffic offloading focused on the study of the \textit{energy-free} interactions between cell association, coverage and throughput. For example, reference \cite{HSJYJSXPJGA12} proposed a biased received power association and studied the average throughput of a downlink channel in a HetNet, but no energy efficiency analysis is involved in this work.  Although the optimal BS intensity problem for achieving high throughput and energy efficiency is studied in \cite{CLJZKBL14}, the impact on the energy efficiency due to cell association is not addressed at all.  

The energy-saving approach for a cellular network by considering dynamic BS operation and cell association together was studied in \cite{ZNYWJGZY10,KSHKYYBZK11}, for example; however, these works only considered the traditional one-tier macro BS model so that its proposed scheme cannot be applied in a multi-tier HetNet. In references \cite{DFCR12,SSHSDJGA13,SSJGA14,BZDGMLH16}, the joint resource allocation, cell activation and traffic offloading problems that originate dynamic traffic variations in a HetNet were investigated and they are approached from the perspective of energy-efficient resource allocation, not from the perspective of making users associate a BS having high energy efficiency. As a result, in general their proposed schemes on cell association and traffic offloading cannot ensure to achieve high link and/or network energy efficiency. Although reference \cite{AMFALACV14} proposed a crucial concept that user association schemes should consider the energy cost in the backhaul communications in order to attain a certain level of energy efficiency, it still does not consider the link energy efficiency of BSs in its proposed context-aware user association scheme.  This motivates the core question delved in this paper -- Can cell association enhance or even optimize the energy efficiency of a BS? Such a question is scarcely studied in prior works on green communications.

\subsection{Contributions}
To investigate the fundamental interplay between cell association schemes and link energy efficiency (i.e., mean spectrum efficiency per unit power consumption at a BS), in this paper we consider a HetNet consisting of $K$ different types of BSs and each particular type of BSs, referred as a tier, are assumed to form a Poisson point process (PPP). We first study the statistical properties of a universal cell association function which is so general that it can cover all pathloss-based cell association schemes in a $K$-tier HetNet and these properties are the basis of deriving all theoretical results in the paper. We then derive the cell load statistics for the power-law cell association function that importantly shows the load balancing status of BSs in every tier and the probability of void BSs  in each tier that do not have any tagged users may not be negligible in a dense HetNet, yet the void BS issue is usually overlooked in almost all prior works on the modeling and analysis in HetNets. The green cell association (GCA) scheme is proposed with a simple and feasible green cell association function that is derived based on the (random) link energy efficiency defined as the spectrum efficiency per unit power consumption at a BS. The GCA function characterizes the ratio of the random received signal power to the power consumption of a BS in the active operating mode.

To assess the link energy efficiency of cell association schemes, we propose the concept of \textit{green coverage probability} that can indicate how likely cell association schemes can attain the desired link energy efficiency.  The tight bounds on the green coverage probability for the scheme GCA are found, which shows the proposed GCA scheme achieves the fundamental maximum limit on the green coverage probability that cannot be surmounted by any other deterministic and/or non-green cell association schemes, such as maximum received power association (MRPA) and nearest BS association (NBA). This maximum limit on the green coverage probability is only achievable whenever  there does not exist random channel impairments such as fading and shadowing or the GCA scheme can completely exploit the channel power gain variations due to those channel impairments.  In other words, there exists an achievability gap in the green coverage probability since cell association schemes in general are unable to complete a new cell association within the short time scale of channel fading variations.

The very tight bounds on the link energy efficiencies for the GCA, MRPA and NBA schemes are all found. Especially, the fundamental maximum limit on the link energy efficiency achieved by GCA is presented in a very neat expression since it is obtained by a special integral technique first devised in this paper. Since the link energy efficiency does not thoroughly reflect the impact from the energy consumption of the void BSs in the dormant operating mode, the network energy efficiency, defined as the mean area spectrum efficiency per unit network power consumption, is proposed to remedy this shortcoming and it is a more realistic and accurate network-wise energy efficiency metric than other energy efficiency metrics defined in the current literatures. All theoretical analyses and numerical results validate that the proposed GCA scheme outperforms non-green cell association schemes such as MRPA and NBA in terms of green coverage probability, link and network energy efficiencies. In addition, GCA also maintains cell load balancing as almost good as NBA that is the best in cell load balancing among all cell association schemes.

\subsection{Paper Organization}
The rest of this paper is organized as follows. In Section \ref{Sec:NetworkModel}, first we introduce the network model of a heterogeneous cellular network consisting of $K$ different types of base stations and the power consumption model of the base stations. Afterwards, we introduce some fundamental results regarding the universal cell association function and cell load statistics. Section \ref{Sec:GCA} presents the analytical results of the green cell association and coverage probability. In Section \ref{Sec:EnergyEfficiency}, the fundamental limits on the energy efficiency of the HetNet with different cell association schemes are investigated. Finally, Section \ref{Sec:Conclusion} concludes our main findings and observations in this paper.

\section{Network Model and Preliminaries}\label{Sec:NetworkModel} 
\subsection{Multi-tier Random Network Model}
Consider a large-scale heterogeneous cellular network (HetNet) on plane $\mathbb{R}^2$ consisting of $K$ different types of base stations (e.g., macrocells, microcells, picocells, etc.) and each specific type of base stations (BSs) is referred as a tier, which is named a $K$-tier HetNet. In this network, all BSs in each tier are assumed to form an independent homogeneous Poisson point processes (PPPs). Specifically, the BSs in the $k$th tier can be described as a marked PPP of intensity $\lambda_k$ and their set $\Phi_k$ can be written as
\begin{align}
\Phi_{k}\defn& \{(B_{k_j}, V_{k_j},H_{k_j}, \Upsilon_{k_j}, \Psi_{k_j}): B_{k_j}\in\mathbb{R}^2, V_{k,j}\in\{0,1\},\nonumber\\ & H_{k_j},\Psi_{k_j},\Upsilon_{k_j}\in\mathbb{R}_{++} \},
\end{align}
where $k\in\mathcal{K}\defn\{1,2,\ldots, K\}$, $B_{k_j}$ denotes the $j$th BS in the $k$th tier and its location, $V_{k_j}$ is a Bernoulli random variable indicating if $B_{k_j}$ is void cell or not, (i.e., if it is associated by any user or not. $V_{k_j}=1$ indicates $B_{k_j}$ is not void and zero otherwise.), $\Psi_{k_j}: \mathbb{R}_+\rightarrow \mathbb{R}_+$ is called the universal cell association (UCA) function\footnote{This ``universal'' cell association function can cover different cell association strategies and it could be random or deterministic depending on the nature of cell association schemes.} used by $B_{k_j}$ and it is bijective, and $H_{k_j}$ is the channel (power) gain from BS $B_{k_j}$ to its serving user and all $H_{k_j}$'s are i.i.d. random variables for same $k$ with all $j\in\mathbb{N}_+$ and independent for all $k\in\mathcal{K}$, $\Upsilon_{k_j}$ stands for the power consumption of BS $B_{k_j}$. Note that all $\Psi_{k_j}$'s are i.i.d. for same $k$ with all $j\in\mathbb{N}_+$ as well and independent for all $k\in\mathcal{K}$ if they contain  random variables. Without loss of generality, \textit{we assume that the first tier in the network is composed of all macro BSs and other $K-1$ tiers are composed of small cell BSs, such as picocells and femtocells}. Also, our following analysis will be based on a typical user located in the origin since the statistics of all location-dependent random parameters seen by this typical user is the same as that seen by all other users in the network due to Slivnyak's theorem\cite{DSWKJM96}. Also, all main variables, symbols and functions used in this paper are summarized in Table \ref{Tab:MathNotation}.

\begin{table}[!t]
  \centering
  \caption{Notation of Main Variables, Symbols and Functions}\label{Tab:MathNotation}
  \begin{tabular}{|c|c|}
  \hline
  Symbol & Definition\\ \hline
   $\lambda_{\mathsf{u}}$ & Intensity of users\\
   $\Phi_{\mathsf{u}}$ &  Homogeneous PPP of Users\\
   $\alpha>2$ & Path loss exponent\\
  $\Phi_k$  & Tier-$k$ Homogeneous PPP \\
  $\lambda_k$ & Intensity of $\Phi_k$\\
  $B_{k_j}$ & $j$th Base station (BS) in the $k$th tier\\
  $H_{k_j}$  & Channel gain of BS $B_{k_j}$\\
  $P_k$ & Transmit power of the tier-$k$ BSs\\
  $\Psi_{k_j}$ & Universal cell association function of BS $B_{k_j}$\\
  $\Upsilon_{k_j}$ & Power consumption of BS $B_{k_j}$\\
  $\Upsilon^{\on}_k$ & Power consumption of active tier-$k$ BSs\\
  $\Upsilon^{\off}_k$ & Power consumption of dormant tier-$k$ BSs\\
  $V_{k_j}\in\{0,1\}$ & one if $B_{k_j}$ is not void and zero otherwise\\
  $p_{k,\emptyset}$ & Void probability of a tier-$k$ BS\\
  $\psi_k$ & Random bias for $\Psi_k(x)=\psi_k x^{-\alpha}$\\
  $\vartheta_k$ & Cell association probability of tier $k$ \\
  $\widetilde{\lambda}_k$ & $\sum_{m=1}^{K}\lambda_m\mathbb{E}[\psi^{2/\alpha}_m]/\mathbb{E}[\psi^{2/\alpha}_k]$\\
  $\zeta_k$ & $\frac{7}{2}\mathbb{E}\left[\psi_k^{\frac{2}{\alpha}}\right]\mathbb{E}\left[\psi_k^{-\frac{2}{\alpha}}\right]$\\
  $\mathsf{L}_k$ & Tier-$k$ cell load\\
  $\rho_{\mathsf{a}}$ & Green coverage probability for association scheme $\mathsf{a}$ \\
  $\delta_{\mathsf{a}}$ & Link energy efficiency for association scheme $\mathsf{a}$ \\
  $\Delta_{\mathsf{a}}$ & Network energy efficiency for association scheme $\mathsf{a}$ \\
  $\eta$ & Threshold of energy efficiency\\
  $g_1\circ g_2(x)$ & Composition function $g_1(g_2(x))$\\
  $f_Z(\cdot) (F_Z(\cdot))$ & pdf (CDF) of random variable $Z$\\
  $\mathcal{L}\{\cdot\}(\mathcal{L}^{-1}\{\cdot\})$ & Laplace (Inverse Laplace) transform operator\\
  $a\gtrapprox b (a\lessapprox b)$ & $b$ is the tight lower (upper) bound on $a$\\
  $\mathbb{R}_+ (\mathbb{R}_{++})$ & Nonnegative real number set (not including zero)\\
  $\mathbb{N} (\mathbb{N}_+)$ & Natural number set (not including zero)\\
  \hline
  \end{tabular}
\end{table}

In this paper, our study is based on the \textit{downlink} transmission scenario and thus all cell association functions are a downlink-based design no matter which user association strategies are adopted\footnote{Since the total power consumption of BSs is much higher than that of mobile handsets and the downlink traffic is usually much higher than the uplink traffic. Accordingly, in this paper we only consider the dowlink network and cell association models because such a downlink formulation can significantly and effectively improve the energy efficiency of BSs in the HetNet.}. Each user associates with a BS that has the maximum cell association function among others. By considering the typical user associating with BS $B^*\in\bigcup_{k\in\mathcal{K}}\Phi_k$, the UCA scheme, without loss of generality, can be written as
\begin{align}\label{Eqn:GenCellAssoScheme}
B^*=\arg\left\{\Psi^*(\|B^*\|)\right\}=\argsup_{B_{k_j}\in\bigcup_{k\in\mathcal{K}}\Phi_k} \Psi_{k_j}(\|B_{k_j}\|), 
\end{align}
where $\|X\|$ is the Euclidean distance from node $X$ to the typical user and $\Psi^*(\cdot)\in\{\Psi_{k_j}(\cdot),\forall k\in\mathcal{K}, j\in\mathbb{N}_+\}$ is the bijective UCA function used by the associated BS  $B^*$. The cell association scheme in \eqref{Eqn:GenCellAssoScheme} can cover any distance-based cell association schemes. For instance, 
users will associate with their nearest BS, i.e., they adopt the (unbiased) nearest BS association  (NBA) scheme with $\Psi_{k_j}(x)=x^{-\alpha}$ only accounting for the path loss of signals where $\alpha>2$ denotes the path loss exponent. Another example is to let BS $B_{k_j}$ have cell association function $\Psi_{k_j}(x)=P_k H_{k_j} x^{-\alpha}$ so that users adopt the (unbiased) maximum received power association (MRPA) scheme to select their BS. In addition, BS $B_{k_j}$ has the following power consumption model dependent on whether it is void or not \cite{GAVGCD11,CHLLCW16}:
\begin{align}\label{Eqn:PowerConsumModel}
\Upsilon_{k_j}=V_{k_j}\Upsilon^{\on}_k+(1-V_{k_j})\Upsilon^{\off}_k,\,\, k\in\mathcal{K},
\end{align}
in which $\Upsilon^{\on}_k\defn P_k^{\on}+\omega_k P_k>1$ stands for the  power consumption while BS $B_{k_j}$ is in the active operating mode and $P_k^{\on}$ is the hardware power consumed by the tier-$k$ BS when the BS is in the active operating mode (i.e., it not void), $P_k$ is the transmit power of the tier-$k$ BS, $\omega_k>1$ is a scaling factor for $P_k$, and $\Upsilon^{\off}_k$ is the power consumed by the tier-$k$ BS in the dormant operating mode (i.e., as it is void.). This power consumption model implicitly contains the power control law for the tier-$k$ BSs, that is, the BSs will be switched from the active operating mode to the dormant operating mode in order to save power provided that they are void. In our prior works in \cite{CTPCHLLCW15,CHLLCW15,CHLLCW1503,CHLLCW16},  we have shown that small cell BSs, unlike marco BSs, have a non-negligible void probability since they are usually densely deployed, and small cell BSs usually can be quickly switched between the active and dormant modes. Note that macro BSs have a deterministic power consumption model as $\Upsilon_{1_j}\approx P_1^{\on}+\omega_1 P_1$ almost surely in that they are hardly void in general.  The power consumption model in \eqref{Eqn:PowerConsumModel} will be used to define the downlink energy efficiency of BSs under any user association schemes in the following section. Later, we will show that the link/network energy efficiency is significantly impacted by how a user associates with its serving BS and we need to introduce some fundamental and general results regarding cell association in the following subsection before validating this point. Those fundamental and general results help us derive the theoretical results of some performance metrics, such as green coverage probability, mean link capacity, link and network energy efficiencies, etc.  

\subsection{Preliminaries of Universal Cell Association}
The following theorem shows the distribution of the value of the user association function corresponding to BS $B^*$, which is important since it can be used to characterize some key performance metrics, such as coverage probability, throughput and energy efficiency, etc.
\begin{theorem}\label{Thm:CDFMaxAssFun}
Suppose the UCA function $\Psi_{k_j}:\mathbb{R}_+\rightarrow\mathbb{R}_+$ are monotonic decreasing, random and independent for all $k\in\mathcal{K}$ and $j\in\mathbb{N}_+$. If random variable $X\in\mathbb{R}_+$ is independent of all $\Psi_{k_j}$'s and the conditional second moment $\mathbb{E}_{\Psi_k}[(\Psi_k^{-1}(X))^2|X]$ exists for all $k\in\mathcal{K}$ where $\Psi_k^{-1}(\cdot)$ denotes the inverse function of $\Psi_{k}(\cdot)$, we have
\begin{align}\label{Eqn:DisMaxAssFun1}
\mathbb{E}\left[F_{\Psi^*(\|B^*\|)}(X)\right]=&\mathbb{E}\bigg\{\exp\bigg(-\pi\sum_{k=1}^K\lambda_k\times\nonumber\\
&\mathbb{E}_{\Psi_k}\left[\left(\Psi^{-1}_k(X)\right)^2\bigg| X\right]\bigg)\bigg\},
\end{align}
where $F_Z(\cdot)$ is the cumulative distribution function (CDF) of random variable $Z$. However, if random variable $X$ and $\Psi_{k_j}$ are correlated, then we have the following
 \begin{align}\label{Eqn:DisMaxAssFun2}
 \mathbb{E}\left[F_{\Psi^*(\|B^*\|)}(X)\right]=\mathbb{E}\left[\exp\left(-\pi\sum_{k=1}^K\lambda_k\left(\Psi^{-1}_k(X)\right)^2\right)\right].
 \end{align} 
\end{theorem}
\begin{IEEEproof}
See Appendix \ref{App:ProofCDFMaxAssFun}.
\end{IEEEproof} 

Due to the generality of the UCA function $\Psi_{k_j}(\cdot)$, the results in Theorem \ref{Thm:CDFMaxAssFun} can be applied in many different user association contexts. For instance, if the UCA function of BS $B_{k_j}$ is the \textit{power-law} (received-signal-power) model from $B_{k_j}$ to the typical user, for example, the MRPA scheme with  $\Psi_{k_j}(\|B_{k_j}\|)=P_kH_{k_j}\|B_{k_j}\|^{-\alpha}$, then $\Psi^{-1}_{k_j}(y)=\left(\frac{P_kH_{k_j}}{y}\right)^{1/\alpha}$ and the complementary cumulative density function (CCDF) of the maximum received power of a user can be found based on  \eqref{Eqn:DisMaxAssFun1} as
\begin{align}\label{Eqn:CCDFMaxRecPower}
&\mathbb{P}\left[\Psi^*(\|B^*\|)\geq \frac{1}{x^{\alpha}}\right]=\mathbb{P}\left[\sup_{B_{k_j}\in\bigcup_{k\in\mathcal{K}}\Phi_k}\frac{P_kH_{k_j}}{\|B_{k_j}\|^{\alpha}}\geq \frac{1}{x^{\alpha}}\right]\nonumber\\
&=1-\exp\left\{-\pi x^2\sum_{k=1}^K\lambda_kP^{\frac{2}{\alpha}}_k\mathbb{E}\left[ H_k^{\frac{2}{\alpha}}\right]\right\},
\end{align}
which indicates this CCDF is the same as the CDF of the path-loss-only received power from the \textit{nearest} point of an equivalent homogeneous PPP of intensity $\sum_{k\in\mathcal{K}}\lambda_kP^{\frac{2}{\alpha}}_k\mathbb{E}\left[H_k^{\frac{2}{\alpha}}\right]$. This fact can be used to calculate the distributions of some performance metrics (such as signal-to-interference ratio (SIR)) if all users associate with their \textit{strongest} BS, i.e., MRPA is adopted. In addition, if $\Psi_{k_j}(x)$ is deterministic and only depends on the path loss, i.e., $\Psi_{k_j}(x)=x^{-\alpha}$, \eqref{Eqn:CCDFMaxRecPower} further reduces to 
\begin{align}
\mathbb{P}\left[\Psi^*(\|B^*\|)\geq \frac{1}{x^{\alpha}}\right]&=\mathbb{P}\left[\sup_{B_{k_j}\in\bigcup_{k\in\mathcal{K}}\Phi_k}\|B_{k_j}\|^{-\alpha}\geq \frac{1}{x^{\alpha}}\right]\nonumber\\
&=1-\exp\left(-\pi x^2\sum_{k=1}^K\lambda_k\right),
\end{align}
which can be used to find the distributions of some performance metrics pertaining to (unbiased) nearest BS association (NBA). 

According to the results in Theorem \ref{Thm:CDFMaxAssFun}, we can derive the probability of associating a BS from any tier and the CDF of the distance between the associated BS and the typical user for deterministic cell association functions as shown in the following theorem.
\begin{theorem}\label{Thm:CDFDisAssBS}
If all UCA functions are random, the probability that the associated BS $B^*$ in \eqref{Eqn:GenCellAssoScheme} from tier $k$, i.e., $\vartheta_k\defn\mathbb{P}[B^*\in\Phi_k]$ is given by
\begin{align}\label{Eqn:CellAssociationProb1}
\vartheta_k=& 2\pi\lambda_k\bigintssss_{0}^{\infty} y\,\mathbb{E}_{\hat{\Psi}_k}\bigg[\exp\bigg(-\pi\sum_{m=1}^K \lambda_m\times\nonumber\\
&\mathbb{E}_{\Psi_m}\left[ \left(\Psi^{-1}_m\circ\hat{\Psi}_k\left(y\right)\right)^2\right]\bigg)\bigg] \dif y,
\end{align}
where $\hat{\Psi}_k(\cdot)$ and $\Psi_k(\cdot)$ are i.i.d. for all $k\in\mathcal{K}$. On the other hand, if all cell association functions are deterministic, the CDF of the distance from BS $B^*$ given in \eqref{Eqn:GenCellAssoScheme} to the typical user is given by
\begin{align}\label{Eqn:CDFDistAssBS}
F_{\|B^*\|}(x) =1-\sum_{k\in\mathcal{K}} \exp\left(-\pi\sum_{m=1}^K\lambda_m\left[\Psi_m^{-1}\circ\Psi_k(x)\right]^2\right) \vartheta_k
\end{align}
in which $\vartheta_k$ in \eqref{Eqn:CellAssociationProb1} reduces to
\begin{align}
\vartheta_k= 2\pi\lambda_k\bigintsss_{0}^{\infty} y\exp\left(-\pi\sum_{m=1}^K \lambda_m \left(\Psi^{-1}_m\circ\Psi_k\left(y\right)\right)^2\right) \dif y.
\end{align}
\end{theorem}
\begin{IEEEproof}
See Appendix \ref{App:ProofCDFDisAssBS}.
\end{IEEEproof}

With the result in \eqref{Eqn:CellAssociationProb1}, we can easily find the 
cell association probability for any cell association scheme. In the case of the aforementioned MRPA scheme with $\Psi_{k_j}(x)=P_kH_{k_j}x^{-\alpha}$, for example, we can find
\begin{align}
\vartheta_k=&\lambda_k \mathbb{E}_{H_k}\bigg[\bigintssss_{0}^{\infty} \exp\bigg(-z \sum_{m=1}^K\lambda_m \left(\frac{P_m}{P_kH_k}\right)^{\frac{2}{\alpha}}\times\nonumber\\
&\mathbb{E}\left[H^{\frac{2}{\alpha}}_m\right] \bigg)\dif z \bigg]=\frac{\lambda_kP^{\frac{2}{\alpha}}_k\mathbb{E}\left[H_k^{\frac{2}{\alpha}}\right]}{\sum_{m=1}^K\lambda_mP^{\frac{2}{\alpha}}_m\mathbb{E}\left[H_m^{\frac{2}{\alpha}}\right]}. \label{Eqn:CellAssProbMRPA}
\end{align}
For the NBA scheme with $\Psi_{k_j}(x)=x^{-\alpha}$, we have the tier-$k$ cell association probability and the CDF of $\|B^*\|$ as follows
\begin{align}
\vartheta_k=\frac{\lambda_k}{\sum_{m=1}^K\lambda_m} \text{ and }F_{\|B^*\|}(x) = 1-e^{-\pi x^2\sum_{k=1}^K\lambda_k},
\end{align}
which coincides with the result in \cite{HSJYJSXPJGA12}. These two user association examples indicate that Theorems \ref{Thm:CDFMaxAssFun} and \ref{Thm:CDFDisAssBS} play a pivotal role in finding the statistics of the signal-to-interference ratio (SIR) and mean spectrum efficiency of a user for any cell association scheme as long as the user association functions of the scheme satisfy the constraints stated in Theorem \ref{Thm:CDFMaxAssFun}.  

\subsection{Cell Load Statistics for Power-Law Cell Association Functions}
In this subsection, we study the distribution of the number of users associated with a BS in a particular tier when the UCA functions in \eqref{Eqn:GenCellAssoScheme} have a power-law form, which is called the cell load statistics that refers to the random number of users carried by a cell. Assume all users in the network also form an independent homogeneous PPP of intensity $\lambda_{\subU}$ denoted by $\Phi_{\subU}\defn\{U_j\in\mathbb{R}^2: j\in\mathbb{N}_+\}$ where $U_j$ represents the $j$th user and its location. 
Let $\mathcal{C}_{k_j}\subset\mathbb{R}^2$ denote the cell region of BS $B_{k_j}$ where all users associating $B_{k_j}$ are located and $\nu(\mathcal{C}_{k_j})$ denote the Lebesgue measure of $\mathcal{C}_{k_j}$. Since the users is a PPP of intensity $\lambda_{\subU}$, the probability that a tier-$k$ BS is associated by $n$ users can be expressed as
\begin{align}
p_{k,n} &\defn\mathbb{P}\left[\Phi_{\subU}(\mathcal{C}_{k_j})=n\right]\nonumber\\
&=\frac{1}{n!}\mathbb{E}\left[(\lambda_{\subU}\nu(\mathcal{C}_k))^ne^{-\lambda_{\subU}\nu(\mathcal{C}_k)}\right],\label{Eqn:DefnNumUserTierkBS}
\end{align}
where $\Phi_{\subU}(\mathcal{C}_{k_j})$ stands for the number of the users associating with BS $B_{k_j}$ and assuming all $\nu(\mathcal{C}_{k_j})$'s are i.i.d. for same $k$. Then the accurate expression of $p_{k,n}$ is provided in the following lemma.
\begin{lemma}\label{Lem:CellLoadStat}
If all users adopt the scheme in \eqref{Eqn:GenCellAssoScheme} to associate their BSs and $\Psi_{k_j}(x)=\psi_{k_j}x^{-\alpha}$ with  $\mathbb{E}\left[\psi^{\frac{2}{\alpha}}_k\right]\mathbb{E}\left[\psi^{-\frac{2}{\alpha}}_k\right]<\infty$ for all $k\in\mathcal{K}$, the probability that there are $n$ users associating with a tier-$k$ BS is given by
\begin{align}
p_{k,n}=\frac{\Gamma(n+\zeta_k)}{n!\,\Gamma(\zeta_k)}\left(\frac{\lambda_{\subU}}{\zeta_k\widetilde{\lambda}_k}\right)^n\left(\frac{\zeta_k\widetilde{\lambda}_k}{\zeta_k\widetilde{\lambda}_k+\lambda_{\subU}}\right)^{n+\zeta_k},\label{Eqn:ProbNumUserTierkBS}
\end{align}
where $\widetilde{\lambda}_k\defn \sum_{m=1}^{K}\lambda_m\mathbb{E}\left[\psi^{\frac{2}{\alpha}}_m\right]\big/\mathbb{E}\left[\psi^{\frac{2}{\alpha}}_k\right]$ and $\zeta_k=\frac{7}{2}\mathbb{E}\left[\psi_k^{\frac{2}{\alpha}}\right]\mathbb{E}\left[\psi_k^{-\frac{2}{\alpha}}\right]$. Then the mean number of the users associating with a tier-$k$ BS, call \textbf{tier-$k$ cell load} denoted by $\mathsf{L}_k$, can be shown as
\begin{align}
\mathsf{L}_k\defn\mathbb{E}\left[\Phi_{\subU}(\mathcal{C}_k)\right]=\frac{\lambda_{\subU}\mathbb{E}\left[\psi^{\frac{2}{\alpha}}_k\right]}{\sum_{m=1}^{K}\lambda_m\mathbb{E}\left[\psi^{\frac{2}{\alpha}}_m\right]}.\label{Eqn:CellLoadTierk}
\end{align}
\end{lemma}
\begin{IEEEproof}
See Appendix \ref{App:ProofCellLoadStat}.
\end{IEEEproof}
 
 The probability in \eqref{Eqn:ProbNumUserTierkBS} not only indicates how the number of users is distributed over different tiers but also reveals whether traffic load is balanced between these $K$ tiers or not, and it highly depends on $\rho_k$ and $\widetilde{\lambda}_k$. For example, the first moment of the number of users associating a tier-$k$ BS given in \eqref{Eqn:CellLoadTierk} indicates the cell loads between tiers are essentially unbalanced unless all $\widetilde{\lambda}_k$'s are the same. In other words, all cell loads will get more balanced if $\lambda_1\mathbb{E}\left[\psi^{\frac{2}{\alpha}}_1\right]\approx \lambda_2\mathbb{E}\left[\psi^{\frac{2}{\alpha}}_2\right]\approx\cdots\approx \lambda_K\mathbb{E}\left[\psi^{\frac{2}{\alpha}}_K\right]$. The parameter $\widetilde{\lambda}_k$ has an interesting physical meaning, that is, it can be interpreted as the equivalent intensity of the network provided that all BSs are composed of the tier-$k$ BSs, and thus $\lambda_k/\widetilde{\lambda}_k$ is exactly the probability that a user associates with a tier-$k$ BS. In addition, \eqref{Eqn:ProbNumUserTierkBS} also importantly shows that it is significantly affected by user association schemes and leads to the void cell probability of a tier-$k$ BS given by
 \begin{align}\label{Eqn:VoidProb}
 \mathbb{P}\left[V_k=0\right]=p_{k,\subVoid}=\left(1+\frac{\lambda_{\subU}}{\zeta_k\widetilde{\lambda}_k}\right)^{-\zeta_k}.
 \end{align}  
 \begin{figure}[!t]
\centering
\includegraphics[width=3.5in,height=2.65in]{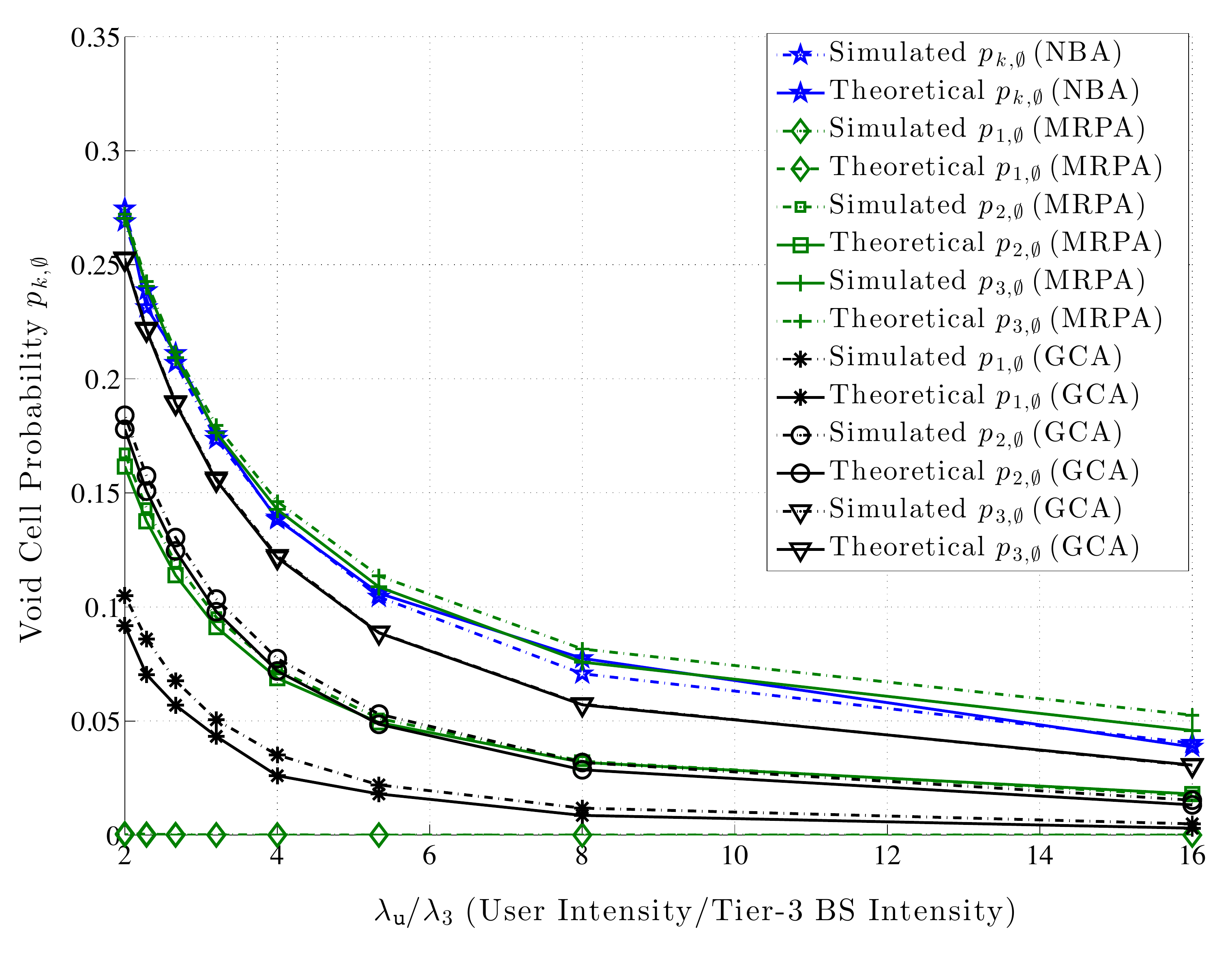}
\caption{The void cell probabilities for a 3-tier HetNet with three cell association schemes -- GCA with $\Psi_{k_j}(x)=P_kH_{k_j}x^{-\alpha}/\Upsilon^{\on}_k$, MRPA with $\Psi_{k_j}(x)=P_kH_{k_j}x^{-\alpha}$ and NBA with $\Psi_{k_j}(x)=x^{-\alpha}$. The network parameters for simulation are $\lambda_1=10^{-6}$ (BS/m$^2$), $\lambda_1:\lambda_2:\lambda_3=1:10:50$, $P_1=40$W, $P_2=1$W, $P_3=0.5$W, $P^{\on}_1=118.7$W, $P^{\on}_2=6.8$, $P^{\on}_3=4.8$W, $\alpha=4$, $\omega_1=2.66$, $\omega_2=4.0$ and $\omega_3=7.5$, exponential (Rayleigh) fading gain with unit mean and log-normal shadowing gain with zero mean and 3dB variance.}
\label{Fig:VoidCellProb}
\end{figure}
Therefore,  the void cell probability for each BS in any tier indeed exists and cannot be negligible if the cell load of a BS is small. To validate this void cell issue, a simulation example for a 3-tier HetNet with three cell schemes -- green cell association (GCA)\footnote{In Section \ref{Sec:GCA}, we will show that this GCA scheme with cell association function $\Psi_{k_j}(x)=P_kH_{k_j}x^{-\alpha}/\Upsilon^{\on}_k$ is able to achieve the maximum energy efficiency among all other cell association schemes.} with $\Psi_{k_j}(x)=P_kH_{k_j}x^{-\alpha}/\Upsilon^{\on}_k$, MRPA with $\Psi_{k_j}(x)=P_kH_{k_j}x^{-\alpha}$ and NBA with $\Psi_{k_j}(x)=x^{-\alpha}$ is illustrated in Fig. \ref{Fig:VoidCellProb}. As can be observed in the figure, all simulated void cell provabilities fairly coincide with their corresponding theoretical values calculated by \eqref{Eqn:VoidProb} and the accuracy of \eqref{Eqn:ProbNumUserTierkBS} is verified. The NBA scheme has a much higher void cell probability than the other two schemes since it does not exploit the fading channel gain variations to reduce its void probabilities\cite{CHLLCW15,CHLLCW16} and thus it allocates/directs the traffic to different kinds of BSs equally likely, whereas the GCA scheme prefers to offload the traffic to the small cells in order to save power and it gives rise to a higher void cell probability for the macro BSs and this induces lower void probabilities happening at small cell BSs. Since the void BSs can be turned off to save power, the void cell probability will apparently impact the analysis of the interference and power consumption in the network and it turns out to be a crucial factor dominating the correctness and accurateness of doing the SIR analysis related to the coverage probability and link capacity.  
 
\section{Green Cell Association and Coverage Probability}\label{Sec:GCA}
Almost all current schemes pertaining to cell association, traffic offloading and cell range expansion are merely developed based on the strength of the received signal power between a user and a BS. The basic spirit of these signal-power-based cell association schemes is to make users associate with a BS providing satisfactory coverage, rate and load balance, however, these schemes fail to characterize the energy usage of BSs and they may attain the cell association objectives of high throughput, coverage and load balance performance at the expense of high power consumption. In a HetNet where densely deployed small cell BSs may induce a large amount of energy consumption, maintaining high data rate and low power consumption is important in order to fulfill the green (energy-efficient) communication goal for such a network.  

The energy efficiency of a BS can be defined as how much the mean spectrum efficiency can be achieved at the cost of the unit power consumption of a BS. Since the downlink throughput of a user depends on the SIR of the user, cell association schemes certainly affect the link throughput as well as the power consumption of a BS. To characterize the fundamental relationship between link spectrum efficiency and power consumption at a BS, the green cell association function is proposed as follows.
\begin{definition}[Green Cell Association Function]
For the $j$th BS in the $k$th tier of the interference-limited HetNet with $K$ tiers, its green cell association (GCA) function is defined as
\begin{align}\label{Eqn:DefnGCAFunction}
\Psi_{k_j}(\|B_{k_j}\|)\defn \log_2\left(1+\frac{P_kH_{k_j}}{I_{k_j}\|B_{k_j}\|^{\alpha}}\right)\bigg/\Upsilon^{\on}_k,
\end{align}
where $I_{k_j}\defn \sum_{B_{m_j}\in\bigcup^K_{m=1}\Phi_k\setminus B_{k_j}}P_mH_{m_j}V_{m_j}\|B_{m_j}\|^{-\alpha}$ denotes the interference evaluated at the origin\footnote{Note that all $I_{k_j}$'s are correlated for all $k\in\mathcal{K}$ and $j\in\mathbb{N}_+$ since they are generated by the same $K$ sets of BSs.} in addition to the signal power from BS $B_{k_j}$, $P_kH_{k_j}/I_{k_j}\|B_{k_j}\|^{\alpha}$ is the SIR from BS $B_{k_j}$ to the typical user and $\Upsilon^{\on}_k$ defined in \eqref{Eqn:PowerConsumModel} is the power consumption of the BS in the active operating mode.
\end{definition}
\noindent Obviously, this GCA function not only characterizes the spectrum efficiency per unit power consumption, but also leverages the SIR and the power consumption at a BS. Thus, the key to improving the energy efficiency of the network is to make a user associate a BS having the largest GCA function for the user. In the following, we will propose a more feasible GCA function for easy implementation that is able to make users associate a BS with high energy efficiency and low computational complexity. 

\subsection{Green Cell Association} 
According to the GCA function defined in \eqref{Eqn:DefnGCAFunction}, we propose the following GCA scheme:
\begin{align}\label{Eqn:DefnGCA}
B^*&=\arg \left\{\Psi^*(\|B^*\|)\right\}\nonumber\\
&\defn\argsup_{B_{k_j}\in\bigcup_{k=1}^K \Phi_k} \Psi_{k_j}(\|B_{k_j}\|),
\end{align}
where $\Psi^*(\cdot)$ is the GCA function corresponding to $B^*$. Such a GCA scheme is somewhat complicate in computation and have a difficulty in implementation if the interference cannot be accurately estimated. Accordingly, the following lemma renders an equivalent low-complexity GCA scheme.
\begin{lemma}\label{Lem:EqnGCA}
The green cell association scheme in \eqref{Eqn:DefnGCA} is equivalently rewritten as the following scheme
\begin{align}\label{Eqn:GCA}
B^*&=\arg\left\{\Psi^*(\|B^*\|)\right\}=\argsup_{B_{k_j}\in\bigcup_{k=1}^K\Phi_k} \frac{P_kH_{k_j}}{\Upsilon^{\on}_k\|B_{k_j}\|^{\alpha}}\nonumber\\
&=\argsup_{B_{k_j}\in\bigcup_{k=1}^K\Phi_k}\Psi_{k_j}(\|B_{k_j}\|) 
\end{align}
in which the GCA function of BS $B_{k_j}$ becomes $\Psi_{k_j}(x)\defn P_kH_{k_j}x^{-\alpha}/\Upsilon^{\on}_k$. 
\end{lemma}
\begin{IEEEproof}
See Appendix \ref{App:ProofEqnGCA}.
\end{IEEEproof}

The GCA scheme in Lemma \ref{Lem:EqnGCA} is essentially a ``biased" MRPA scheme in which the cell association function needs to characterize the ratio of the received signal power to the power consumption of an active BS in order to achieve high energy efficiency. In fact, the important implication of the GCA scheme in Lemma \ref{Lem:EqnGCA} is that users tend to associate with a BS that has a high transmit power faction of the power consumption, which intuitively makes sense because having high transmit power (i.e., high SIR and spectrum efficiency) and low hardware power consumption is the key to attaining high energy efficiency. For tractability and simplicity in the following analysis, we will use the GCA scheme in \eqref{Eqn:GCA} instead of that in \eqref{Eqn:DefnGCA}. The probability in \eqref{Eqn:ProbNumUserTierkBS} for the GCA scheme, for instance, can be easily found by substituting $\mathbb{E}\left[\psi_k^{\frac{2}{\alpha}}\right]=\left(\frac{P_k}{\Upsilon^{\on}_k}\right)^{\frac{2}{\alpha}}\mathbb{E}[H_k^{\frac{2}{\alpha}}]$ into $\zeta_k$ and $\widetilde{\lambda}_k$, i.e.,
\begin{align}\label{Eqn:GCARhok}
\zeta_k=\frac{7}{2}\mathbb{E}\left[H_k^{\frac{2}{\alpha}}\right]\mathbb{E}\left[H_k^{-\frac{2}{\alpha}}\right]
\end{align}
and
\begin{align}\label{Eqn:GCAEqnIntensityTierk}
\widetilde{\lambda}_k=\dfrac{\sum_{m\in\mathcal{K}}\lambda_m\left(P_m/\Upsilon^{\on}_m\right)^{\frac{2}{\alpha}}\mathbb{E}\left[H_m^{\frac{2}{\alpha}}\right]}{\left(P_k/\Upsilon^{\on}_k\right)^{\frac{2}{\alpha}}\mathbb{E}\left[H_k^{\frac{2}{\alpha}}\right]}=\frac{\lambda_k}{\vartheta_k},
\end{align}
which also immediately gives the probability $\vartheta_k$ of associating a tier-$k$ BS and can be used to find the tier-$k$ cell load in \eqref{Eqn:CellLoadTierk} for the GCA scheme. 

\subsection{Green Coverage Probability}\label{Subsec:GreenCoverProb}
We have realized that user association schemes indeed significantly influence the energy efficiency of a BS.  To quantitatively assess the fundamental interplay between energy efficiency and user association, the green coverage probability of an associated BS, based on the energy efficiency concept, is defined as follows. 
\begin{definition}[Green Coverage Probability] We define the green coverage probability as the performance metric of indicating how likely a BS can maintain the desired energy
efficiency under any particular cell association scheme. As a result, the green coverage probability for cell association scheme ``$\mathsf{a}$'', denoted by $\rho_{\mathsf{a}}$, can be mathematically expressed as
\begin{align}
\rho_{\mathsf{a}}(\eta)\defn  \mathbb{P}\left[\sum_{k=1}^{K}\log_2\left(1+\frac{P_kH_k}{I^*\|B^*\|^{\alpha}}\right)\frac{\mathds{1}(B^*\in\Phi_k)}{\Upsilon^{\on}_k}\geq\eta\right],\label{Eqn:DefnGreenCoverage}
\end{align}
where subscript $\mathsf{a}\in\{\gc,\mr,\nb,\ldots\}$\footnote{Throughout this paper, subscripts $\gc$, $\mr$ and $\nb$ denote the GCA, MRPA, and NBA schemes, respectively.}, $\eta>0$ is the threshold of the desired energy efficiency\footnote{Actually, the green coverage probability is another form of the traditional SIR-based coverage probability. Thus, the following analytical results of the green coverage probability can be transformed to the traditional coverage probability by replacing $2^{\eta\Upsilon^{\on}_k-1}$ with the desired SIR threshold.}, $I^*\defn \sum_{B_{m_j}\in\bigcup^K_{m=1}\Phi_k\setminus B^*}P_mH_{m_j}V_{m_j}\|B_{m_j}\|^{-\alpha}$ denotes the interference received by the typical user, and $\mathds{1}(\mathcal{E})$ is the indicator function that is equal to one if event $\mathcal{E}$ is true and zero otherwise.  
\end{definition}
\noindent According to the definition of the green coverage probability, a BS is said \textit{in the outage of energy efficiency} if it cannot reach its desired energy efficiency while serving users. For the GCA scheme in \eqref{Eqn:GCA}, its green coverage probability can be shown in the following theorem.
\begin{theorem}\label{Thm:GreenCoverageProbGCA}
For the GCA scheme in \eqref{Eqn:GCA}, the tight lower bound on the green coverage probability can be found and shown in \eqref{Eqn:GreenCoverProbGCA} 
\begin{figure*}
\begin{align}
\rho_{\gc}(\eta)\gtrapprox \sum_{k=1}^{K} \vartheta_k\bigintsss_{0}^{1}\mathcal{L}^{-1}\left\{\left[1+\sum_{m=1}^{K}\hbar\left(s\frac{\Upsilon^{\on}_m(2^{\eta\Upsilon^{\on}_k}-1)}{\Upsilon^{\on}_k},\frac{2}{\alpha}\right)(1-p_{m,\subVoid})\vartheta_m\right]^{-1}\right\}(\tau) \dif\tau, \label{Eqn:GreenCoverProbGCA}
\end{align}
\hrulefill
\end{figure*}
where $a\gtrapprox b$ ($a\lessapprox b$) means $b$ is the tight lower (upper) bound on $a$, $\hbar(x,y)\defn x^y\left[\Gamma\left(1-y\right)+y\Gamma\left(-y,x\right)\right]-1$ for $y\in(0,1)$, $\vartheta_k=\lambda_k(P_k/\Upsilon^{\on}_k)^{\frac{2}{\alpha}}\mathbb{E}\left[H^{\frac{2}{\alpha}}_k\right]/\sum_{m=1}^{K}\lambda_m(P_m/\Upsilon^{\on}_m)^{\frac{2}{\alpha}}\mathbb{E}\left[H^{\frac{2}{\alpha}}_m\right]$, $p_{m,\subVoid}$ is given in \eqref{Eqn:VoidProb} with $\zeta_k$ given in \eqref{Eqn:GCARhok}, and $\widetilde{\lambda}_k$ given in \eqref{Eqn:GCAEqnIntensityTierk} and $\mathcal{L}^{-1}\{G(s)\}(\tau)$ is the inverse Laplace transform operator of function $G(s)$. 
\end{theorem}
\begin{IEEEproof}
See Appendix \ref{App:ProofGreenCoverageProbGCA}.
\end{IEEEproof}

The tight lower bound in \eqref{Eqn:GreenCoverProbGCA} is derived based on the fact stated in the proof of Theorem \ref{Thm:GreenCoverageProbGCA} that the distribution of non-void cells still behaves very similarly to PPPs even though they theoretically do not form PPPs any more due to the presence of the location correlations between the non-void BSs. Thus, as the user intensity goes to infinity (i.e., the void cell probability goes to zero), the interference increases so that $\rho_{\gc}$ reduces and eventually converges to its lower bound without void cell probabilities, that is,
\begin{align}
\lim_{\lambda_{\subU}\rightarrow\infty}\rho_{\gc}(\eta)=&\sum_{k=1}^{K}\vartheta_k\int_{0}^{1}\mathcal{L}^{-1}\bigg\{\bigg[1+\sum_{m=1}^{K}\vartheta_m\times\nonumber\\
&\hbar\left(s\frac{\Upsilon^{\on}_m(2^{\eta\Upsilon^{\on}_k}-1)}{\Upsilon^{\on}_k},\frac{2}{\alpha}\right)\bigg]^{-1}\bigg\}\left(\tau\right)\dif\tau, \label{Eqn:LowestLimitGreenCovProbGCA}
\end{align}
\textit{which is the lowest limit $\rho_{\gc}$ can achieve and does not depend on the intensities of users and BSs}. Also, this lowest limit is the green coverage probability if void cells are overlooked in the interference model and this implies the green coverage probability without considering the void cell impact is actually underestimated. On the other hand, we should have $\lim_{\lambda_k\rightarrow\infty} \rho_{\gc}=1$ since the void cell probabilities go to unity and that leads to an infinitely large SIR. Deploying more BSs indeed benefits the green coverage probability, but it suffers the problem of diminishing marginal returns on the green coverage gain. In general, the tight lower bound on $\rho_{\gc}$ in \eqref{Eqn:GreenCoverProbGCA} cannot be obtained in closed-form since the inverse Laplace transform is usually intractable. Nonetheless, we can resort to numerical Laplace transform methods to find the exact tight lower bound, or by applying $\hbar(x,\frac{2}{\alpha})\gtrapprox x^{\frac{2}{\alpha}}\Gamma(1-\frac{2}{\alpha})-1$ for $x>0$ in \eqref{Eqn:GreenCoverProbGCA}, a tight upper bound on $\rho_{\gc}$ can be obtained as shown in \eqref{Eqn:ApproxBoundsGreenCovProbGCA},
\begin{figure*}
\begin{align}\label{Eqn:ApproxBoundsGreenCovProbGCA}
\rho_{\gc}(\eta)\lessapprox&\sum_{k=1}^{K} \vartheta_k\bigintss_{0}^{1}\mathcal{L}^{-1}\left\{\dfrac{1/\sum_{m=1}^{K}(\Upsilon^{\on}_m)^{\frac{2}{\alpha}}(1-p_{m,\subVoid})\vartheta_m}{\Gamma\left(1-\frac{2}{\alpha}\right)\left(\frac{s(2^{\eta\Upsilon^{\on}_k}-1)}{\Upsilon^{\on}_k}\right)^{\frac{2}{\alpha}}+\frac{\sum_{m=1}^{K}p_{m,\subVoid}\vartheta_m}{\sum_{m=1}^{K}(\Upsilon^{\on}_m)^{\frac{2}{\alpha}}(1-p_{m,\subVoid})\vartheta_m}}\right\}(\tau)\dif\tau,
\end{align}
\hrulefill
\end{figure*}
which has a closed form for $\alpha=4$ (see the Lapalce transform table in\cite{MAIAS72}) and it also has the following closed-form result if $\lambda_{\subU}\rightarrow\infty$:
\begin{align}\label{Eqn:TightUppBoundGreenCoverProbGCA}
\lim_{\lambda_{\subU}\rightarrow\infty}\rho_{\gc}(\eta)\lessapprox \frac{\alpha\sin(2\pi/\alpha)\sum_{k=1}^{K}\vartheta_k\left(\Upsilon^{\on}_k/2^{\eta\Upsilon^{\on}_k}-1\right)^{\frac{2}{\alpha}}}{2\pi\sum_{m=1}^{K}\vartheta_m\left(\Upsilon^{\on}_m\right)^{\frac{2}{\alpha}}}.
\end{align}

It is worth pointing out that \textit{the green coverage probability in \eqref{Eqn:GreenCoverProbGCA} is the fundamental maximum limit of the green coverage probability in that the GCA scheme in \eqref{Eqn:GCA} captures all random channel variations to exploit the ``multi-BS diversity gain" while performing cell association, and this limit cannot be surmounted by any other non-green cell association schemes}. This limit is only achievable whenever there are no random channel impairments such as fading and shadowing, in other words, it may not be practically achievable provided that channel power gains experience small-scaling fast fading since users may not be able to finish another new cell association action within a very short channel coherence time. As a result, if the channel power variations cannot be exploited by green cell association, the green coverage probability in \eqref{Eqn:GreenCoverProbGCA} will reduce, as shown in the following theorem.
\begin{theorem}\label{Thm:GreenCoveProbGCA2}
Suppose all channels in the network undergo independent Rayleigh fading and shadowing and the channel gain from BS $B_{k_j}$ to the typical user can be specifically expressed as $H_{k_j}=H^f_{k_j}\cdot Q_{k_j}$ where fading gain $H^f_{k_j}$ is an exponential random variable with unit mean for all $k_j$ and $Q_{k_j}$ is the random channel gain due to shadowing. If the GCA scheme only can exploit the channel shadowing gain, the GCA function in \eqref{Eqn:GCA} for BS $B_{k_j}$ becomes $\Psi_{k_j}(x)=P_kQ_{k_j}x^{-\alpha}/\Upsilon^{\on}_k$ and the tight lower bound on the green coverage probability can be found in closed-form as
\begin{align}
\rho_{\gc}\gtrapprox&\sum_{k=1}^{K} \vartheta_k \bigg(1+\sum_{m=1}^{K}(1-p_{m,\subVoid})\vartheta_m\times\nonumber\\
&\ell\left[\frac{\Upsilon^{\on}_m(2^{\eta\Upsilon^{on}_k}-1)}{\Upsilon^{\on}_k},\frac{2}{\alpha}\right]\bigg)^{-1}  , \label{Eqn:GreenCoveProbGCA2}
\end{align}
where $\vartheta_k=\lambda_k\left(\frac{P_k}{\Upsilon^{\on}_k}\right)^{\frac{2}{\alpha}}\mathbb{E}\left[Q^{\frac{2}{\alpha}}_k\right]/\sum_{m=1}^{K}\lambda_m\left(\frac{P_m}{\Upsilon^{\on}_m}\right)^{\frac{2}{\alpha}}\mathbb{E}\left[Q^{\frac{2}{\alpha}}_m\right]$, $p_{m,\subVoid}$ is already given in \eqref{Eqn:VoidProb} with $\widetilde{\lambda}_k=\frac{\lambda_k}{\vartheta_k}$ and $\zeta_k=\frac{7}{2}\mathbb{E}\left[Q^{\frac{2}{\alpha}}_k\right]\mathbb{E}\left[Q^{-\frac{2}{\alpha}}_k\right]$, and $\ell(\cdot,\cdot)$ is defined as
\begin{align}
\ell(x,z)&=x^z\left(\frac{\pi z}{\sin(\pi z)}-\int_{0}^{x^{-z}}\frac{\dif t}{1+t^{\frac{1}{z}}}\right)\nonumber\\
&=x^z\left(\frac{\pi z}{\sin(\pi z)}+\int_{0}^{x^{-z}}\frac{t^{\frac{1}{z}}\dif t}{1+t^{\frac{1}{z}}}\right)-1.
\end{align}
\end{theorem}
\begin{IEEEproof}
See Appendix \ref{App:ProofGreenCoveProbGCA2}.
\end{IEEEproof}
\noindent Note that $\rho_{\gc}$ in \eqref{Eqn:GreenCoveProbGCA2} should be smaller than that in \eqref{Eqn:GreenCoverProbGCA} because it does not exploit the multi-BS diversity induced by Rayleigh fading via cell association schemes. \textit{In other words, $\rho_{\gc}$ in \eqref{Eqn:GreenCoveProbGCA2} is much more achievable than that in \eqref{Eqn:GreenCoverProbGCA} if channel fading is unavoidable in practice so that there exists a fundamental gap between them}, thereby we can further deduce that the green coverage probability will reduce more if channel shadowing variations cannot be exploited while performing cell association. Also, as the user intensity goes to infinity $\rho_{\gc}$ in \eqref{Eqn:GreenCoveProbGCA2} will also decrease and converge to the lowest limit given by
\begin{align}
\lim_{\lambda_{\subU}\rightarrow\infty}\rho_{\gc}=\sum_{k=1}^{K} \frac{\vartheta_k}{1+\sum_{m=1}^{K}\ell\left(\frac{\Upsilon^{\on}_m(2^{\eta \Upsilon^{\on}_k}-1)}{\Upsilon^{\on}_k},\frac{2}{\alpha}\right)\vartheta_m},\label{Eqn:LowLimitGCAGreenCoverProbFading}
\end{align}
which can be viewed as the ``more achievable'' lowest limit of the green coverage probability. 

In addition to the green coverage probability for the GCA scheme found in above, the green coverage probabilities for other cell association schemes can also be found based on the similar approaches to deriving the green coverage probability of the GCA scheme. For example, the tight lower and upper bounds on the green coverage probability of the MRPA scheme with $\Psi_{k_j}(x)=P_kH_{k_j}x^{-\alpha}$ can be easily found by slightly modifying the results in \eqref{Eqn:GreenCoverProbGCA} and \eqref{Eqn:ApproxBoundsGreenCovProbGCA}, respectively, and they are given in \eqref{Eqn:MRPAGreenCovProb1}
\begin{figure*}
\begin{align}\label{Eqn:MRPAGreenCovProb1}
\rho_{\mr}(\eta)
\begin{cases}
\gtrapprox \sum\limits_{k=1}^{K}\vartheta_k\bigintss_{0}^{1}\mathcal{L}^{-1}\left\{\left(1+\hbar\left(s(2^{\eta\Upsilon^{\on}_k}-1),\frac{2}{\alpha}\right)\sum\limits_{m=1}^{K}(1-p_{m,\subVoid})\vartheta_m\right)^{-1}\right\}\left(\tau\right)\dif\tau \\
\lessapprox \sum\limits_{k=1}^{K}\vartheta_k\bigintss_{0}^{1}\mathcal{L}^{-1}\left\{\left(\left[s(2^{\eta\Upsilon^{\on}_k}-1)\right]^{\frac{2}{\alpha}}\Gamma(1-\frac{2}{\alpha})\sum\limits_{m=1}^{K}(1-p_{m,\subVoid})\vartheta_m+\sum\limits_{m=1}^{K}p_{m,\subVoid}\vartheta_m\right)^{-1}\right\}\left(\tau\right)\dif\tau,
\end{cases}
\end{align}
\hrulefill
\end{figure*}
where $\vartheta_k=\lambda_kP_k\mathbb{E}\left[H^{\frac{2}{\alpha}}_k\right]/\sum_{m=1}^{K}=\lambda_mP_m\mathbb{E}\left[H^{\frac{2}{\alpha}}_m\right]$ and their asymptotic results for $\lambda_{\subU}\rightarrow\infty$ are given in \eqref{Eqn:MRPAGreenCovProb2}.
\begin{figure*}
\begin{align}\label{Eqn:MRPAGreenCovProb2}
\lim_{\lambda_{\subU}\rightarrow\infty} \rho_{\mr}(\eta) \begin{cases}
= \sum\limits_{k=1}^{K}\vartheta_k\bigintss_{0}^{1}\mathcal{L}^{-1}\left\{\left(1+\hbar\left(s(2^{\eta \Upsilon^{\on}_k}-1),\frac{2}{\alpha}\right)\right)^{-1}\right\}\left(\tau\right)\dif\tau\\
\lessapprox \frac{\alpha\sin(2\pi/\alpha)}{2\pi}\sum\limits_{k=1}^K \vartheta_k(2^{\eta\Upsilon^{\on}_k}-1)^{-\frac{2}{\alpha}}
\end{cases}.
\end{align}
\hrulefill
\end{figure*}
Intuitively, $\rho_{\mr}$ is definitely smaller than $\rho_{\gc}$ since $\rho_{\gc}$ is the maximum green coverage probability that can be achieved by cell association. This fact can also be inferred by comparing \eqref{Eqn:MRPAGreenCovProb1} with \eqref{Eqn:GreenCoverProbGCA}: the integral and $\vartheta_k$ in \eqref{Eqn:GreenCoverProbGCA} both decrease as $P^{\on}_k$ (or $\Upsilon^{\on}_k$) increases, but the integral in \eqref{Eqn:MRPAGreenCovProb1} decreases and $\vartheta_k$  remains unchanged as $P^{\on}_k$ increases. This reveals that $\rho_{\gc}$ is more optimal than $\rho_{\mr}$ since $\rho_{\gc}$ gets more contribution from the BSs with less hardware power consumption, which means $\rho_{\gc}$ is ``greener'' than $\rho_{\mr}$ from the energy-efficiency perspective. Another example is the NBA scheme with $\Psi_{k_j}(x)=x^{-\alpha}$ and the tight lower and upper bounds on the green coverage probability of the NBA scheme, based on Proposition 2 in \cite{CHLLCW16} and the proof of Theorem \ref{Thm:GreenCoverageProbGCA}, can be derived as shown in \eqref{Eqn:NBAGreenCoverProb1}
\begin{figure*}
\begin{align}\label{Eqn:NBAGreenCoverProb1}
\rho_{\nb}\begin{cases}
\gtrapprox\sum\limits_{k=1}^{K}   \vartheta_k\bigintss_{0}^{1}\mathcal{L}^{-1}\left\{\mathbb{E}_{H_k}\left[\left(1+(1-p_{\subVoid})\sum\limits_{m=1}^{K}\mathbb{E}_{H_m}\left[\hbar\left(s_kH_mP_m,\frac{2}{\alpha}\right)\right]\vartheta_m\right)^{-1}\right]\right\}(\tau) \dif\tau\\
\lessapprox \sum\limits_{k=1}^{K} \vartheta_k \bigintss_{0}^{1}\mathcal{L}^{-1}\left\{\mathbb{E}_{H_k}\left[\left(s_k^{\frac{2}{\alpha}}\Gamma(1-\frac{2}{\alpha})(1-p_{\subVoid})\sum\limits_{m=1}^{K}P^{\frac{2}{\alpha}}_m\mathbb{E}\left[H^{\frac{2}{\alpha}}_m\right]\vartheta_m+p_{\subVoid}\right)^{-1}\right]\right\}(\tau) \dif\tau
\end{cases},
\end{align}
\hrulefill 
\end{figure*}
where $p_{\subVoid}=p_{1,\subVoid}=\cdots=p_{K,\subVoid}$, $s_k\defn \frac{s\left(2^{\eta\Upsilon^{\on}_k}-1\right)}{H_kP_k}$ and $\vartheta_k=\lambda_k/\sum_{m=1}^{K}\lambda_m$ and we also have $\lim_{\lambda_{\subU}\rightarrow\infty}\rho_{\nb}$ as shown in \eqref{Eqn:NBAGreenCoverProb2}. 
\begin{figure*}
\begin{align}\label{Eqn:NBAGreenCoverProb2}
\lim_{\lambda_{\subU}\rightarrow\infty}\rho_{\nb}\begin{cases}
=\sum\limits_{k=1}^{K}   \vartheta_k\bigintss_{0}^{1}\mathcal{L}^{-1}\left\{\mathbb{E}_{H_k}\left[\left(1+\sum\limits_{m=1}^{K}\mathbb{E}_{H_m}\left[\hbar\left(s_kH_mP_m,\frac{2}{\alpha}\right)\right]\vartheta_m\right)^{-1}\right]\right\}(\tau) \dif\tau\\
\lessapprox \frac{\alpha\sin(2\pi/\alpha)\sum_{k=1}^{K}\vartheta_k\left(P_k/2^{\eta\Upsilon^{\on}_k}-1\right)^{\frac{2}{\alpha}}\mathbb{E}\left[H^{\frac{2}{\alpha}}_k\right] }{2\pi\sum_{m=1}^{K}\vartheta_mP^{\frac{2}{\alpha}}_m\mathbb{E}\left[H^{\frac{2}{\alpha}}_m\right]}
\end{cases}.
\end{align}
\hrulefill
\end{figure*}
Note that \eqref{Eqn:NBAGreenCoverProb1} and \eqref{Eqn:NBAGreenCoverProb2} are valid for any channel models and they can be further simplified to a neater expression if all channels undergo Rayleigh fading. Since NBA does not exploit any channel variations at all, its green coverage probability must be smaller than that of GCA and might be also smaller than that of MRPA if the power consumption of BSs does not significantly degrade the link energy efficiency. As a result, we can conclude that $\rho_{gc}(\eta)>\max\{\rho_{\nb}(\eta),\rho_{\mr}(\eta)\}$
for a given $\eta$. This fact will be demonstrated in the following subsection of numerical results. 

\subsection{Numerical Results}\label{Subsec:SimulationGreenCoverProb}
\begin{table}[!t]
\centering
\caption{Network Parameters for Simulation\cite{GAVGCD11}}\label{Tab:SimPara}
\begin{tabular}{|c|c|c|c|}
\hline Parameter $\setminus$ BS (Tier \#)& Macro (1) & Picocell (2) & Femtocell (3)\\ \hline
Transmit Power $P_k$ (W) & 40 & 1.0  & 0.05 \\ \hline
Active Power $P^{\on}_k$ (W) & $118.7$ & $6.8$ & $4.8$ \\ \hline 
Dormant Power $P^{\off}_k$ (W) & $93$ & $4.3$ & $2.9$ \\ \hline 
Power Scaling Factor $\omega_k$ & $5.32$ & $4.0$ & $7.5$ \\ \hline
Intensity $\lambda_k$ (BSs/$m^2$) & $1\times 10^{-6}$ & $10\lambda_1$ & $50\lambda_1$ \\ \hline 
Threshold $\eta$ (bps/joule) & \multicolumn{3}{c|}{0.1} \\ \hline  
 Gain $H_{k_j}=H^{f}_{k_j}Q_{k_j}$ & \multicolumn{3}{c|}{$H^f_{k_j}\sim\text{Exp}(1,1)$, $Q_{k_j} \sim \ln\mathcal{N}$(0, 3dB)} \\ \hline
Pathloss Exponent $\alpha$ &\multicolumn{3}{c|}{4}\\ \hline 
\end{tabular} 
\end{table}

\begin{figure}[!t]
\centering
\includegraphics[width=3.7in,height=2.65in]{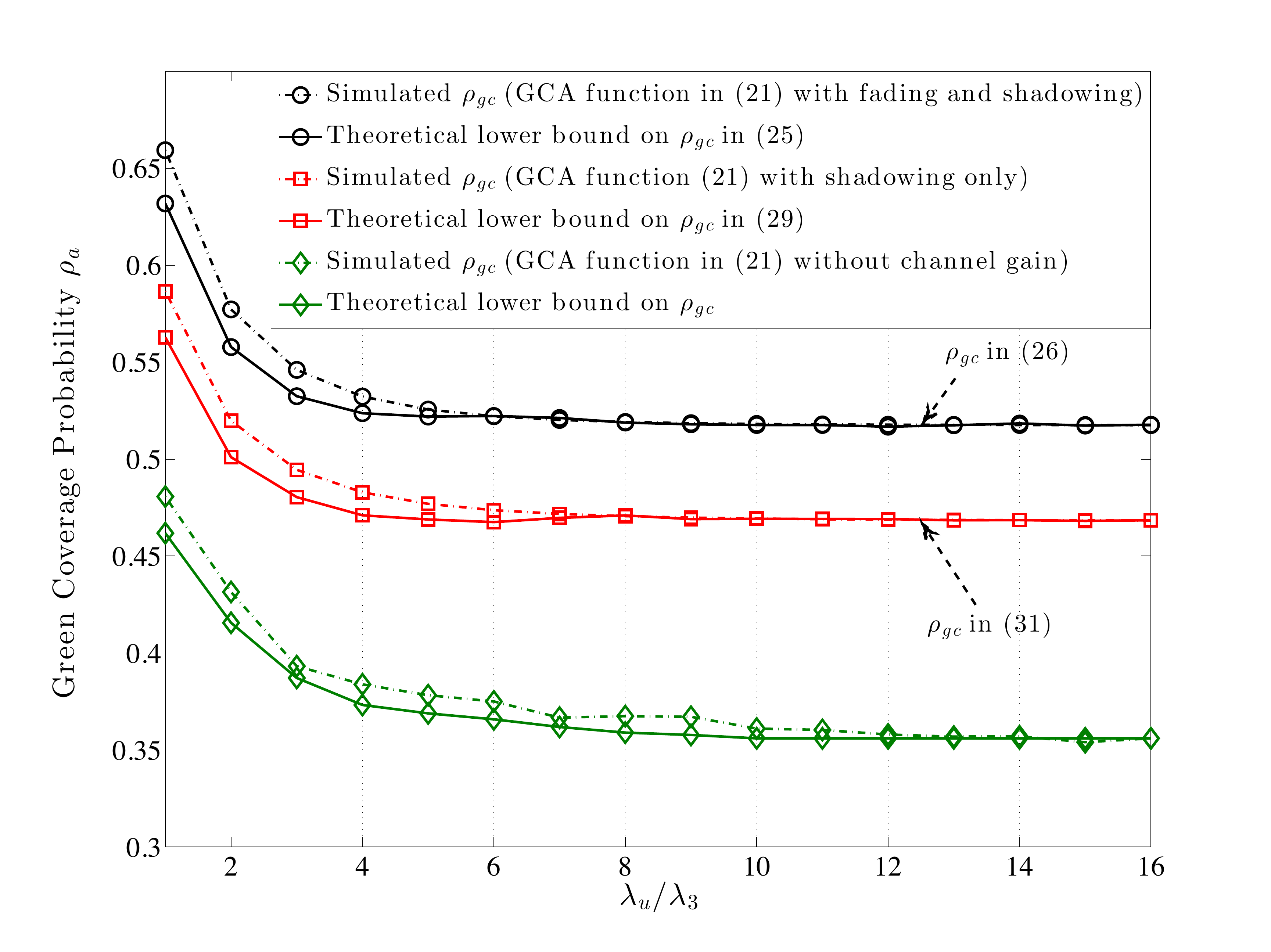}
\caption{Simulation results of the green coverage probabilities of the GCA scheme that captures different channel gains. }
\label{Fig:GreenCoverProb1}
\end{figure}

\begin{figure}[!t]
\centering
\includegraphics[width=3.7in,height=2.5in]{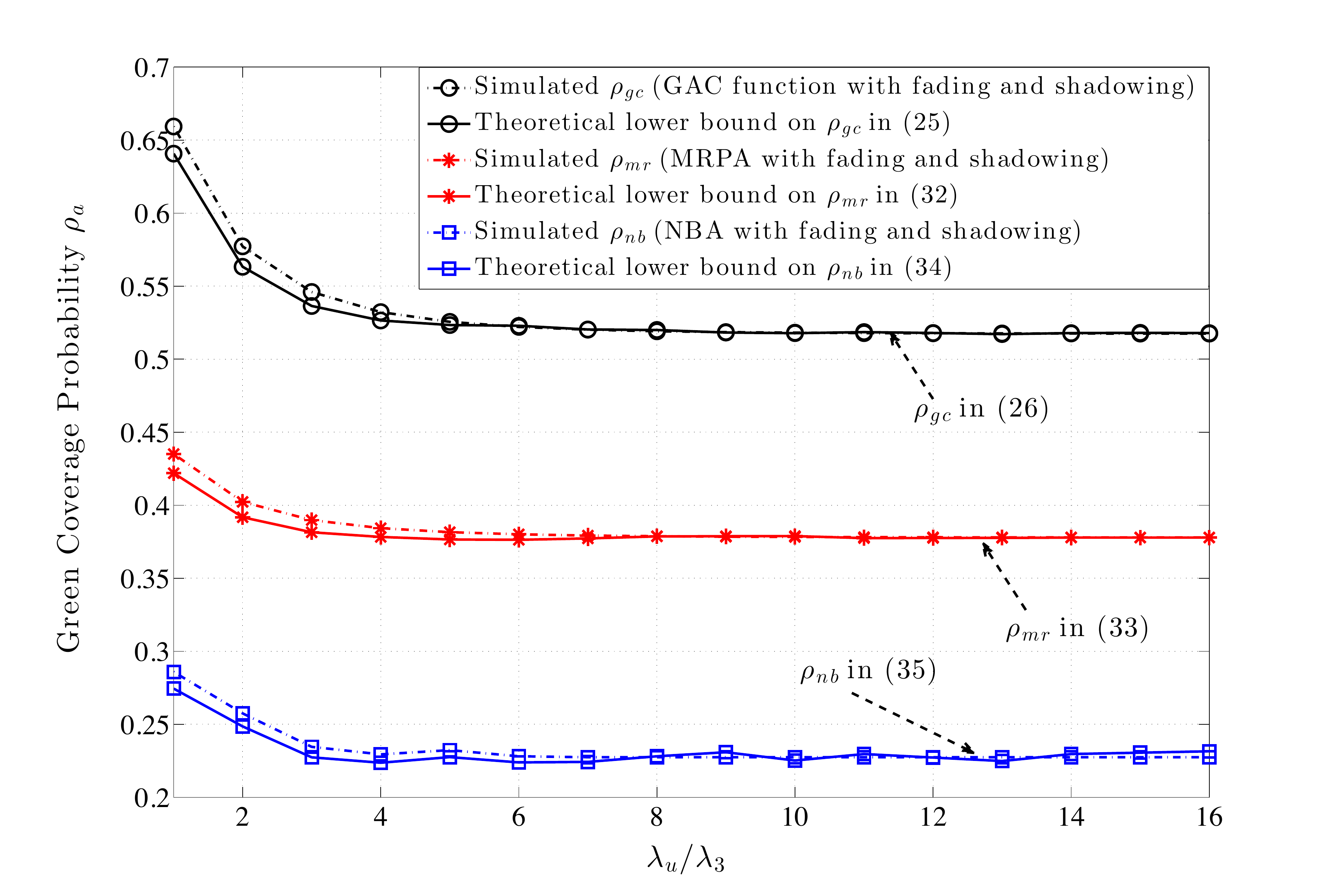}
\caption{Simulation results of the green coverage probabilities of the GCA, MRPA and NBA schemes.}
\label{Fig:GreenCoverProb2}
\end{figure}

\begin{figure}[!t]
\centering
\includegraphics[width=3.5in,height=2.65in]{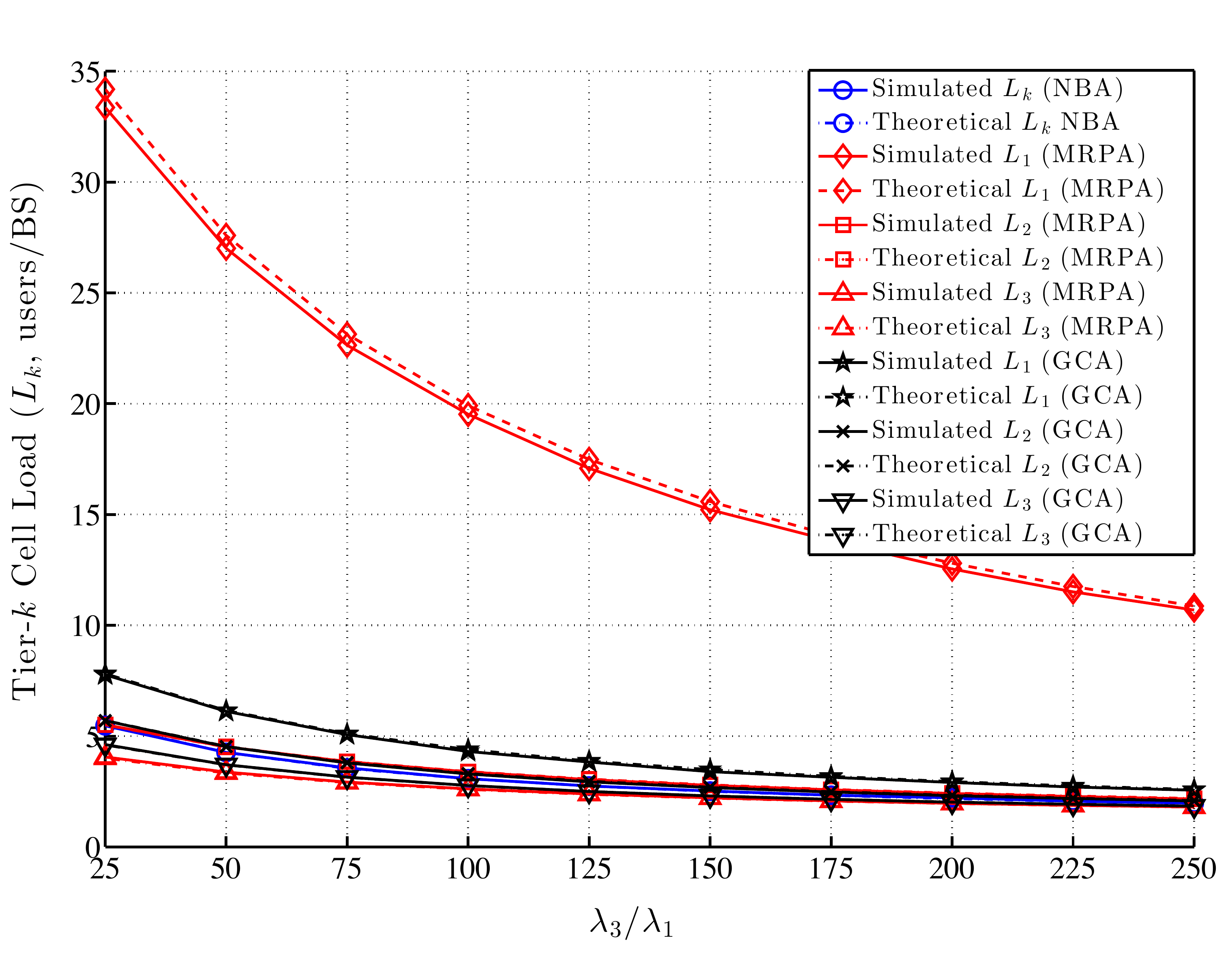}
\caption{Simulation results of the cell load of each tier for GCA, MRPA and NBA. For this simulation, the user intensity is 370 users/km$^2$ and the cell loads are presented by changing the intensity of the BSs in the third tier normalized by the intensity of the BSs in the first tier.}
\label{Fig:CellLoad}
\end{figure}

In this subsection, we first provide some simulation results in a 3-tier HetNet consisting of macrocell, picocell and femtocell BSs to numerically verify the green coverage probabilities derived in Section \ref{Subsec:GreenCoverProb} for the GCA, MRPA and NBA schemes.  Assume all channels undergo Rayleigh fading and log-normal shadowing and all network parameters for simulation are listed in Table \ref{Tab:SimPara}. In Fig. \ref{Fig:GreenCoverProb1}, we show the green coverage probabilities of the GCA scheme that captures three different channel fading situations. As expected, the green coverage probability in \eqref{Eqn:GreenCoverProbGCA} that exploits the Rayleigh fading and log-normal shadowing gains in the channels is much higher than the other two probabilities. All the three simulated green coverage probabilities are tightly lower bounded by their corresponding theoretical lower bounds derived in Theorems \ref{Thm:GreenCoverageProbGCA} and \ref{Thm:GreenCoveProbGCA2} and they eventually converge to different constants as the user intensity goes to infinity. In the case of GCA with fading and shadowing, for example, $\rho_{\gc}$ finally converges to 0.58 found by \eqref{Eqn:ApproxBoundsGreenCovProbGCA} that is slightly smaller than 0.601 found by \eqref{Eqn:TightUppBoundGreenCoverProbGCA}, as expected. The differences between these three green coverage probabilities, as we have emphasized in above,  suggest the fundamental gaps existing in the green coverage performance and we will very likely have to endure a data transmission without the anticipated energy efficiency if the GCA scheme does not contain channel state information (see how random fading and shadowing channel gains severely degrade the green coverage probability in Fig. \ref{Fig:GreenCoverProb1}.). The simulation results of the green coverage probabilities for the GCA, MRPA and NBA schemes are shown in Fig. \ref{Fig:GreenCoverProb2}. Unsurprisingly, GCA outperforms the other two schemes and MRPA outperforms NBA. The tightness of all the derived lower bounds on the green coverage probabilities of the three schemes in \eqref{Eqn:GreenCoverProbGCA}, \eqref{Eqn:MRPAGreenCovProb1} and \eqref{Eqn:NBAGreenCoverProb1} is also confirmed.  

Another good aspect of evaluating the performance of a cell association scheme is to observe whether the scheme is able to balance the cell loads between the BSs at different tiers or not. Having load balancing between BSs is important since it can make users acquire more resource and better service.  For any cell association scheme, its theoretical tier-$k$ cell load can be calculated by \eqref{Eqn:CellLoadTierk} as long as its tier-$k$ cell association function is known. The simulation results of the cell load of each tier for the GCA, MRPA and NBA schemes are illustrated in Fig \ref{Fig:CellLoad}.  According to Fig \ref{Fig:CellLoad}, GCA and NBA both have a very similar performance in cell load balancing and this suggests GCA can balance the cell loads very well since NBA treats different BSs equally and makes the cell loads in different tiers equal, i.e., it completely balances the cell loads.  MRPA is intrinsically unable to balance the cell loads well because it makes users favor the BSs having large transmit power.

\section{Energy Efficiency Analysis}\label{Sec:EnergyEfficiency}
The green coverage probabilities for different cell association schemes have been studied in Section \ref{Sec:GCA} and they are the basis of calculating the energy efficiency of communication links. The link energy efficiency can directly display how much energy efficiency a BS in the active operating mode can achieve on average. Nonetheless, it cannot offer a good evaluation for the energy efficiency of the entire network since there could be some void BSs that still consume some power as shown in \eqref{Eqn:PowerConsumModel} even though they are in the dormant operating mode. In this section, first we investigate how to analyze the link energy efficiency and then propose a new metric of the network energy efficiency to realistically characterize the mean area spectrum efficiency at the cost of the power consumption per unit area in the network.  

\subsection{Analysis of the Link Energy Efficiency}
According to the green coverage probability in \eqref{Eqn:DefnGreenCoverage}, the link energy efficiency of cell association scheme $\mathsf{a}$ can be principally defined as shown in the following
\begin{align}
\delta_{\mathsf{a}} \defn \int_{0}^{\infty} \rho_{\mathsf{a}}(\eta)\dif\eta=\sum_{k=1}^{K}\frac{\mathds{C}_{\mathsf{a},k}}{\Upsilon^{\on}_k}\vartheta_k, \label{Eqn:DefnLinkEngEffic}
\end{align} 
where subscript $\mathsf{a}\in\{\gc,\mr,\nb,\cdots\}$, $\mathds{C}_{\mathsf{a},k}$ is the mean spectrum efficiency given that $B^*\in\Phi_k$ is found by cell association scheme $\mathsf{a}$ and it is given by
\begin{align}
\mathds{C}_{\mathsf{a},k}=\mathbb{E}\left[\log_2\left(1+\frac{P_kH_k}{I^*\|B^*\|^{\alpha}}\right)\bigg| B^*\in\Phi_k\right]. \label{Eqn:MeanLinkCapacity}
\end{align}
For the GCA scheme, its link energy efficiency has a tight lower bound given in the following theorem.
\begin{theorem}\label{Thm:GCAEnergyEff}
Consider all users adopt the GCA scheme in \eqref{Eqn:DefnGCA} to associate their BS in the network. The tight lower bound on the link energy efficiency (bits/Hz/Joule) of the GCA scheme is shown as
\begin{align}\label{Eqn:GCALinkEnergyEff}
\delta_{\gc}\gtrapprox\frac{1}{\ln 2}\sum_{k=1}^{K}\frac{\vartheta_k}{\Upsilon^{\on}_k}\underline{\mathds{C}_{\gc,k}}, 
\end{align}
where $\underline{\mathds{C}_{\gc,k}}$ is the lower bound on $\mathds{C}_{\gc,k}$ and it is given by
\begin{align}
\underline{\mathds{C}_{\gc,k}}\defn \bigintsss_{0^+}^{\infty} \frac{(1-e^{-s})\dif s}{s\left(1+\sum_{m=1}^{K}\vartheta_m(1-p_{m,\subVoid})\hbar\left(\frac{s\Upsilon^{\on}_m}{\Upsilon^{\on}_k},\frac{2}{\alpha}\right) \right)}
\end{align}
in which the exact results of $\vartheta_k$ and $\hbar(\cdot,\cdot)$  can be referred to Theorem \ref{Thm:GreenCoverageProbGCA}.
\end{theorem}
\begin{IEEEproof}
See Appendix \ref{App:ProofGCAEnergyEff}.
\end{IEEEproof}
\noindent \textit{The link energy efficiency in \eqref{Eqn:GCALinkEnergyEff} achieved by the GCA scheme, like the case of $\rho_{\gc}$, is the fundamental maximum limit on the link energy efficiency in the network and it cannot be surmounted by any other non-green cell association schemes}. Applying $\hbar(x,\frac{2}{\alpha})\gtrapprox x^{\frac{2}{\alpha}}\Gamma(1-\frac{2}{\alpha})-1$ in \eqref{Eqn:GCALinkEnergyEff} gives the tight upper bound on $\delta_{\gc}$ as shown in \eqref{Eqn:GCALinkEnergyEffUpp}.
\begin{figure*}
\begin{align}\label{Eqn:GCALinkEnergyEffUpp}
\delta_{\gc}\lessapprox \frac{1}{\ln 2}\sum_{k=1}^{K} \bigintsss_{0^+}^{\infty} \frac{\vartheta_k\left(\Upsilon^{\on}_k\right)^{\frac{2}{\alpha}-1}(1-e^{-s})\dif s}{s\left(s^{\frac{2}{\alpha}}\sum_{m=1}^{K}(1-p_{m,\subVoid})\left(\Upsilon^{\on}_m\right)^{\frac{2}{\alpha}}\vartheta_m+\sum_{m=1}^{K}p_{m,\subVoid}\vartheta_m\right)}. 
\end{align}
\hrulefill
\end{figure*}
As all the BS intensities go to infinity, the link energy efficiency goes to infinity as well because the void probabilities go to unity and the SIR goes to infinity as well, that is, $\lim_{\lambda_k\rightarrow\infty}\rho_{\gc} =\infty$, which means \textit{deploying more BSs can increase the link energy efficiency for a given user intensity}. Furthermore, as the user intensity goes to infinity $\delta_{\gc}$ and its tight bound will converge to their lowest limits given in \eqref{Eqn:LinkEELimitGC}
\begin{figure*}
\begin{align}\label{Eqn:LinkEELimitGC}
\lim_{\lambda_{\subU}\rightarrow\infty}\delta_{\gc}=\begin{cases}
\frac{1}{\ln 2}\sum_{k=1}^{K}\frac{\vartheta_k}{\Upsilon^{\on}_k} \int_{0^+}^{\infty} \left(1+\sum_{m=1}^{K}\hbar\left(\frac{s\Upsilon^{\on}_m}{\Upsilon^{\on}_k},\frac{2}{\alpha}\right) \vartheta_m\right)^{-1}\frac{(1-e^{-s})}{s}\dif s\\
\lessapprox \frac{1}{\ln 2}\frac{\sum_{k=1}^{K}\vartheta_k(\Upsilon^{\on}_k)^{\frac{2}{\alpha}-1} }{\left(\sum_{m=1}^{K}\left(\Upsilon^{\on}_m\right)^{\frac{2}{\alpha}}\vartheta_m\right)} \int_{0^+}^{\infty}\frac{(1-e^{-s})}{s^{1+\frac{2}{\alpha}}}\dif s
\end{cases}.
\end{align}
\hrulefill
\end{figure*}
which do not depend on BS intensities and this means \textit{deploying more BSs hardly improves the link energy efficiency whenever the user intensity is extremely larger than the BS intensities}. This lowest limit also represents the link energy efficiency for the context where void cells are overlooked in the interference model,  as the similar case explained in the green coverage probability. Be aware that $\delta_{\gc}$ in \eqref{Eqn:GCALinkEnergyEff} can only be achieved whenever channels do not suffer any random channel impairments such as fading and/or shadowing, whereas it is fairly difficult to be achieved in a fading environment since the GCA scheme hardly exploits the fast fading channel variations in general, like the green coverage probability in \eqref{Eqn:GreenCoverProbGCA}.  The more achievable link energy efficiency, given in the following theorem, is obtained by making the GCA scheme exploit the shadowing variations in the channels. 
\begin{theorem}
Consider the same fading channel gain model and the same GCA function specified in Theorem \ref{Thm:GreenCoveProbGCA2}. The tight lower bound on $\delta_{\gc}$ is given by
\begin{align}\label{Eqn:GCALinkEnergyEff2}
\delta_{\gc}\gtrapprox& \frac{1}{\ln 2}\sum_{k=1}^{K}\frac{\vartheta_k}{\Upsilon^{\on}_k}\bigintsss_{0}^{\infty}\bigg[(1+s)\times\nonumber\\ &\left(1+\sum_{m=1}^{K}\ell\left(s\frac{\Upsilon^{\on}_m}{\Upsilon^{\on}_k},\frac{2}{\alpha}\right)(1-p_{m,\subVoid})\vartheta_m\right)\bigg]^{-1}\dif s,
\end{align}
where $\vartheta_k$ and $p_{m,\subVoid}$ are defined in Theorem  \ref{Thm:GreenCoveProbGCA2}. If the user intensity goes to infinity, then all void cell probabilities vanish and it follows that 
\begin{align}
\lim_{\lambda_{\subU}\rightarrow\infty} \delta_{\gc} = &\frac{1}{\ln 2}\sum_{k=1}^{K}\frac{\vartheta_k}{\Upsilon^{\on}_k} \bigintsss_{0}^{\infty}\bigg[(1+s)\times\nonumber\\ &\left(1+\sum_{m=1}^{K}\ell\left(s\frac{\Upsilon^{\on}_m}{\Upsilon^{\on}_k},\frac{2}{\alpha}\right)\vartheta_m\right)\bigg]^{-1}\dif s. \label{Eqn:GCALinkEnergyEffLowLimit}
\end{align}
\end{theorem}
\begin{IEEEproof}
Since the tight lower bound on the green coverage probability has already been found in \eqref{Eqn:GreenCoveProbGCA2}, we can integrate it with respect to $\eta$ to find the tight lower bound on $\delta_{\gc}$ as shown in the following
\begin{align*}
\delta_{\gc}=&\int_{0}^{\infty}\rho_{\gc}(\eta)\dif \eta \gtrapprox \sum_{k=1}^{K}\vartheta_k \bigintsss_{0}^{\infty} \bigg(1+\sum_{m=1}^{K}(1-p_{m,\subVoid})\vartheta_m\times\\
&\hspace{1.25in}\ell\left(\frac{\Upsilon^{\on}_m(2^{\eta\Upsilon^{on}_k}-1)}{\Upsilon^{\on}_k},\frac{2}{\alpha}\right)\bigg)^{-1}\dif\eta.
\end{align*}
Then replacing variable $\eta$ in the integral with $\log_2(1+s)/\Upsilon^{\on}_k$ leads to \eqref{Eqn:GCALinkEnergyEff2} and letting $p_{m,\subVoid}$ go to zero yields \eqref{Eqn:GCALinkEnergyEffLowLimit}.
\end{IEEEproof}

Without considering the fading effect in the GCA scheme, $\delta_{\gc}$ in \eqref{Eqn:GCALinkEnergyEff2} must be smaller than that in \eqref{Eqn:GCALinkEnergyEff}, like the aforementioned discussion about the green coverage probability, the fundamental gap also exists in the link energy efficiency. As for the MRPA scheme, its tight bounds on link energy efficiency that can be easily inferred from \eqref{Eqn:GCALinkEnergyEff} and \eqref{Eqn:GCALinkEnergyEffUpp} are
\begin{align}\label{Eqn:MRPALinkEnergyEff}
\delta_{\mr}\begin{cases}
\gtrapprox\frac{1}{\ln 2} \int_{0^+}^{\infty} \frac{(1-e^{-s})\dif s}{s\left(1+\hbar\left(s,\frac{2}{\alpha}\right)\sum\limits_{m=1}^{K}\vartheta_m(1-p_{m,\subVoid}) \right)}\sum\limits_{k=1}^{K}\frac{\vartheta_k}{\Upsilon^{\on}_k}\\
\lessapprox \frac{1}{\ln 2} \int_{0^+}^{\infty} \frac{(1-e^{-s})\dif s}{s^{1+\frac{2}{\alpha}}}\sum\limits_{k=1}^{K}\frac{\vartheta_k}{\Upsilon^{\on}_k}
\end{cases}
\end{align}
where $\vartheta_k$ is the same as that in \eqref{Eqn:MRPAGreenCovProb1} and
\begin{align}
\lim_{\lambda_{\subU}\rightarrow\infty} \delta_{\mr}&=\frac{1}{\ln 2} \int_{0^+}^{\infty} \frac{(1-e^{-s})\dif s}{s\left(1+\hbar\left(s,\frac{2}{\alpha}\right) \right)}\sum_{k=1}^{K}\frac{\vartheta_k}{\Upsilon^{\on}_k}\nonumber\\
&\lessapprox \frac{1}{\ln 2}\int_{0^+}^{\infty} \frac{(1-e^{-s})\dif s}{s^{1+\frac{2}{\alpha}}}\sum_{k=1}^{K}\frac{\vartheta_k}{\Upsilon^{\on}_k} .
\end{align}
For the NBA scheme, its tight bounds on the link energy efficiency  can be found by \eqref{Eqn:NBAGreenCoverProb1} and they are shown in \eqref{Eqn:UppLowBoundsLinkEENBA} 
\begin{figure*}
\begin{align}\label{Eqn:UppLowBoundsLinkEENBA}
\delta_{\nb}\begin{cases}
\gtrapprox \frac{1}{\ln 2}\sum\limits_{k=1}^{K}\frac{\vartheta_k}{\Upsilon^{\on}_k} \int_{0}^{\infty}(1+s)^{-1}\mathbb{E}_{H_k}\left[ \left(1+(1-p_{\subVoid})\sum\limits_{m=1}^{K}\mathbb{E}_{H_m}\left[\hbar\left(s\frac{H_mP_m}{H_kP_k},\frac{2}{\alpha}\right)\right]\vartheta_m\right)^{-1}\right]\dif s\\
\lessapprox \frac{1}{\ln 2}\sum\limits_{k=1}^{K}\frac{\vartheta_k}{\Upsilon^{\on}_k} \int_{0}^{\infty}(1+s)^{-1}\mathbb{E}_{H_k}\left[ \left(s_k^{\frac{2}{\alpha}}\Gamma(1-\frac{2}{\alpha})(1-p_{\subVoid})\sum\limits_{m=1}^{K}P^{\frac{2}{\alpha}}_m\mathbb{E}\left[H^{\frac{2}{\alpha}}_m\right]\vartheta_m+p_{\subVoid}\right)^{-1}\right]\dif s
\end{cases},
\end{align}
\hrulefill
\end{figure*}
in which $\vartheta_k$ is the same as that in \eqref{Eqn:NBAGreenCoverProb1} so that we have $\lim_{\lambda_{\subU}\rightarrow\infty} \delta_{\nb}$ shown in \eqref{Eqn:LinkEELimitNBA}. 
\begin{figure*}
\begin{align}\label{Eqn:LinkEELimitNBA}
\lim_{\lambda_{\subU}\rightarrow\infty} \delta_{\nb}\begin{cases}
= \frac{1}{\ln 2}\sum_{k=1}^{K}\frac{\vartheta_k}{\Upsilon^{\on}_k} \int_{0}^{\infty}(1+s)^{-1}\mathbb{E}_{H_k}\left[ \left(1+\sum_{m=1}^{K}\mathbb{E}_{H_m}\left[\hbar\left(s\frac{H_mP_m}{H_kP_k},\frac{2}{\alpha}\right)\right]\vartheta_m\right)^{-1}\right]\dif s\\
\lessapprox \frac{\sum_{k=1}^{K}\vartheta_k(\Upsilon^{\on}_k)^{-1}P_k^{\frac{2}{\alpha}}\mathbb{E}\left[H_k^{\frac{2}{\alpha}}\right]}{(\ln 2)\Gamma(1-\frac{2}{\alpha})\sum_{m=1}^{K}P^{\frac{2}{\alpha}}_m\mathbb{E}\left[H^{\frac{2}{\alpha}}_m\right]\vartheta_m} \int_{0}^{\infty}\frac{\dif s}{s^{\frac{2}{\alpha}}(1+s)}
\end{cases}.
\end{align}
\hrulefill
\end{figure*}
Similarly, $\delta_{\gc}$ also must be greater than $\delta_{\nb}$, but the ordering between $\delta_{\mr}$ and $\delta_{\nb}$ cannot be absolutely discerned. Thus, we also can conclude $\delta_{\gc}>\max\{\delta_{\mr},\delta_{\nb}\}$, which will be validated in the section of numerical results.

\subsection{Analysis of the Network Energy Efficiency}
The link energy efficiency has a drawback, that is, it fails to characterize the power consumed by those void BSs and cannot provide an accurate overall index of how the entire HetNet efficiently uses its own energy. Hence, we propose the following network energy efficiency to remedy this drawback.
\begin{definition}[Network Energy Efficiency]
 The network energy efficiency for cell association scheme ``$\mathsf{a}$'' is defined as the ratio of the mean area spectrum efficiency of the network to the total power consumption of the network per unit area and its expression can be written as \footnote{Note that the network energy efficiency defined in this paper is essentially based on the concept of ``the link throughput intensity", which may be different from the network energy efficiency defined based on the entire network throughput and power consumption.}
\begin{align}\label{Eqn:DefnNetworkEnergyEff}
\Delta_{\mathsf{a}} &\defn \frac{\sum_{k=1}^{K}\lambda_k(1-p_{k,\subVoid})\mathds{C}_{\mathsf{a},k}}{\sum_{m=1}^{K}\lambda_m\mathbb{E}[\Upsilon_m]}\nonumber\\
&=\frac{\sum_{k=1}^{K}\lambda_k(1-p_{k,\subVoid})\mathds{C}_{\mathsf{a},k}}{\sum_{m=1}^{K}\lambda_m[(1-p_{m,\subVoid})\Upsilon^{\on}_m+p_{m,\subVoid}\Upsilon^{\off}_m]},
\end{align}
where $\Upsilon_m$ is defined in \eqref{Eqn:PowerConsumModel} and $\mathds{C}_{\mathsf{a},k}$ is the mean spectrum efficiency of an associated BS in tier $k$ and already defined in \eqref{Eqn:MeanLinkCapacity}.
\end{definition}
\noindent The salient feature of this network energy efficiency, compared with the network-wise energy efficiency defined in the prior works (such as \cite{EMDKCS15} and \cite{LXXGCHWFLFR13}), is to characterize the void cell issue that impacts network capacity and energy consumption of BSs so that this network energy efficiency is a more accurate metric of evaluating how much area spectrum efficiency is achieved at the cost of the unit energy consumed in the network per unit area.  

The network energy efficiencies of the GCA, MRPA and NBA schemes can be directly inferred from the link energy efficiencies $\delta_{\gc}$, $\delta_{\mr}$ and $\delta_{\nb}$, respectively. For example,  according to the proof of Theorem \ref{Thm:GCAEnergyEff}, the tight lower bound on $\mathds{C}_{\gc,k}$ is obtained, i.e., $\underline{\mathds{C}_{\gc,k}}$, and substituting it into \eqref{Eqn:DefnNetworkEnergyEff} readily yields the network energy efficiency of the GCA scheme given by
\begin{align}\label{Eqn:GCANetworkEnergyEff}
\Delta_{\gc}\gtrapprox\frac{\sum_{k=1}^{K}\lambda_k(1-p_{k,\subVoid})\underline{\mathds{C}_{\gc,k}}}{(\ln 2)\sum_{m=1}^{K}\lambda_m[(1-p_{m,\subVoid})\Upsilon^{\on}_m+p_{m,\subVoid}\Upsilon^{\off}_m]},
\end{align}
and note that $p_{k,\subVoid}$ and $\vartheta_k$ are the same as those given in  Theorem \ref{Thm:GreenCoverageProbGCA}. As the user intensity goes to infinity, $\Delta_{\gc}$ converges to the following lowest limit
\begin{align}
\lim_{\lambda_{\subU}\rightarrow\infty} \Delta_{\gc}=\frac{\sum_{k=1}^{K}\lambda_k \underline{\mathds{C}^{\infty}_{\gc,k}}}{(\ln 2)\sum_{m=1}^{K}\lambda_m\Upsilon^{\on}_m},
\end{align}
where $\underline{\mathds{C}^{\infty}_{\gc,k}}\defn \lim_{\lambda_{\subU}\rightarrow \infty}\underline{\mathds{C}_{\gc,k}}$, 
and $\Delta_{\gc}$ goes to zero as all BS intensities go to infinity, i.e., $\lim_{\lambda_k\rightarrow\infty}\Delta_{\gc}=0$, because the total power consumed in the network goes to infinity whereas the mean area spectrum efficiency of the network converges to a constant. Due to the limited space, the tight bounds on the network energy efficiencies of the MRPA and NBA schemes are not specified here, however, we can expect that they have the same convergence properties as those in the case of the GCA scheme.

\subsection{Numerical Results}
\begin{figure}
\centering
\includegraphics[width=3.75in,height=2.65in]{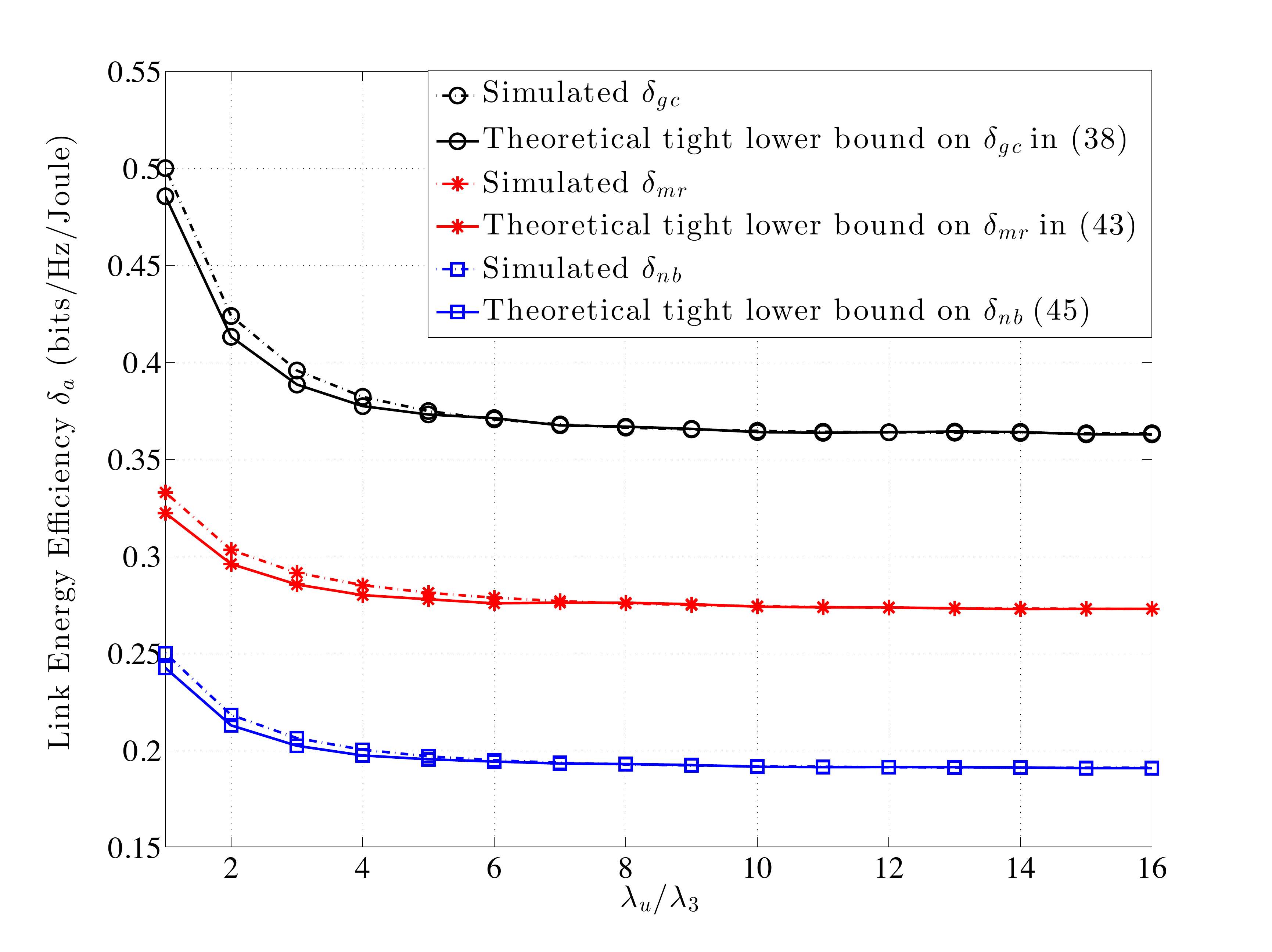}
\caption{Simulation results for the link energy efficiencies of the GCA, MRPA and NBA schemes.}
\label{Fig:LinkEE}
\end{figure}
\begin{figure}
\centering
\includegraphics[width=3.75in,height=2.65in]{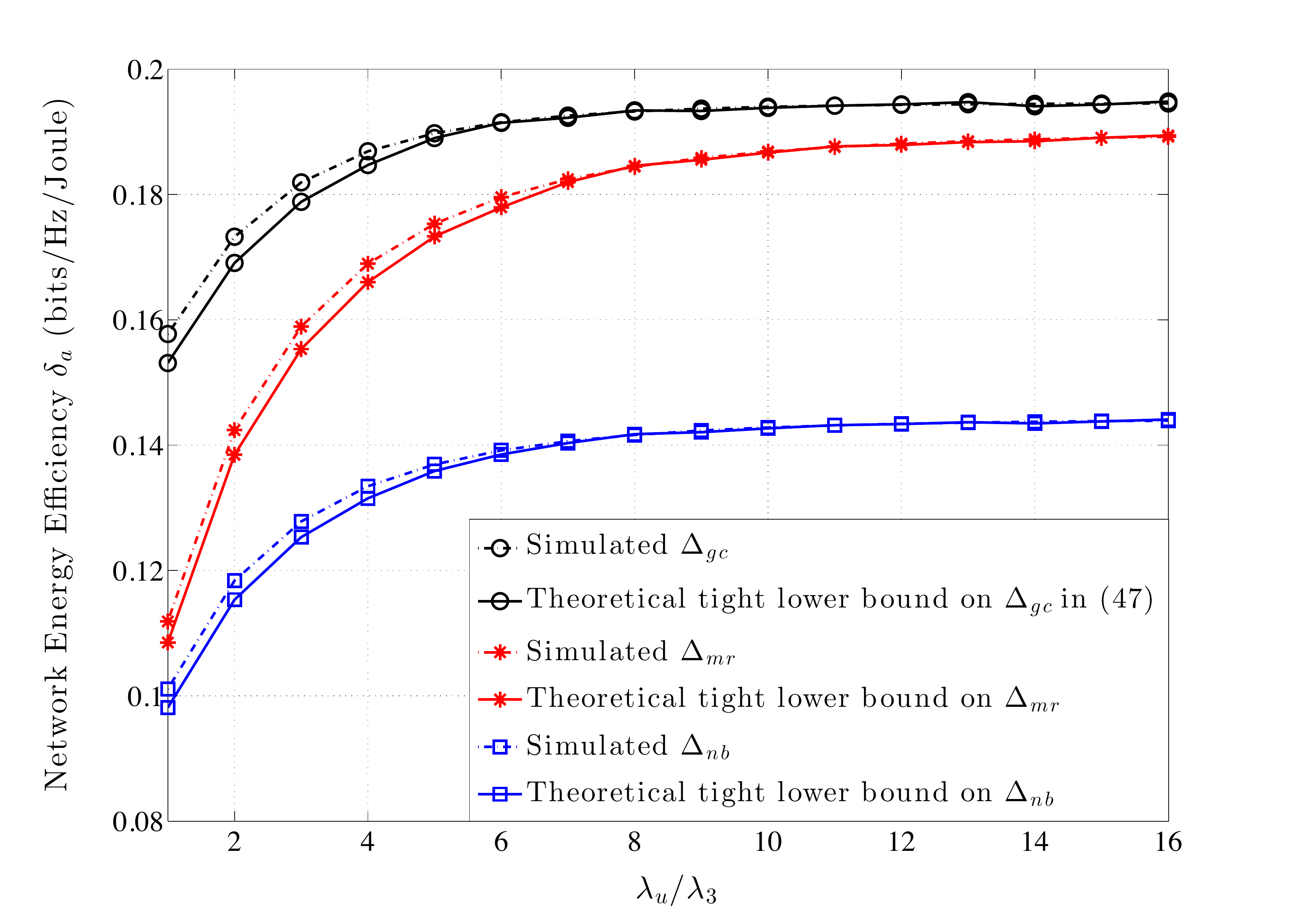}
\caption{Simulation results for the network energy efficiencies of the GCA, MRPA and NBA schemes.}
\label{Fig:NetworkEE}
\end{figure}

In this subsection, two numerical simulations are provided to validate the previous derivations of the link and network energy efficiencies and our findings. All simulation assumptions are the same as those in Section \ref{Subsec:SimulationGreenCoverProb} and the network parameters for simulation are also the same as those listed in Table \ref{Tab:SimPara}. The simulation results of the link energy efficiencies are shown in Fig. \ref{Fig:LinkEE} and GCA is also the best among the three schemes and its link energy efficiency poses the fundamental maximum limit on the link energy efficiency, as expected. All link energy efficiencies reduce and eventually converge to some constants as the user intensity goes to infinity, which means BSs are getting to lose their ``greenness'' as more users join in the network. Fig. \ref{Fig:NetworkEE} shows the simulation results of the network energy efficiencies that are quiet different from those of the link energy efficiencies. In the low region of $\lambda_{\subU}/\lambda_3$, GCA has the lowest network power consumption among the three schemes since it makes a considerable portion of the macro BSs be in the dormant mode if compared with the other two schemes and saves a lot of power (see the void probabilities shown in Fig. \ref{Fig:VoidCellProb}). Since GCA also has a higher link energy efficiency that MRPA and NBA, its network efficiency is significantly higher than those of MRPA and NBA due to low network power consumption. When $\lambda_{\subU}/\lambda_3$ increases, all network energy efficiencies increase and approach to constants since all mean area spectrum efficiencies increase faster than all network power consumptions of different cell association schemes that increase up to a constant.

\section{Conclusion}\label{Sec:Conclusion}
In this paper, we investigate the fundamental limits on the energy efficiency in a HetNet from the cell association perspective. We propose the green cell association scheme that can leverage the received (random) signal power and the power consumption of an active BS and also propose the green coverage probability to evaluate how likely a cell association scheme can achieve the desired link energy efficiency: the more green coverage probability, the better link energy efficiency. The network energy efficiency is also proposed to realistically assess the area spectrum efficiency per unit network power consumption contributed by all active and dormant BSs. Our important findings are first to derive all tight bounds on the green coverage probability and link energy efficiency of the GCA scheme and then show that they are the fundamental maximum limits on those achieved by other cell association schemes, such as MRPA and NBA. Namely, the proposed GCA scheme outperforms other schemes in terms of the link and network energy efficiencies.  

\appendix[Proofs of Lemmas and Theorems]

\subsection{Proof of Theorem \ref{Thm:CDFMaxAssFun}}\label{App:ProofCDFMaxAssFun}
First consider random variable $X$ are independent of all $\Psi_{k_j}$'s. In this case, $\mathbb{E}\left[F_{\Psi^*(\|B^*\|)}(X)\right]$ is equal to $\mathbb{P}\left[\sup_{B_{k_j}\in\bigcup_{k=1}^K\Phi_k} \Psi_{k_j}(\|B_{k_j}\|)\leq X\right]$ given by
\begin{align*}
&\mathbb{P}\left[\sup_{B_{k_j}\in\bigcup_{k=1}^K\Phi_k} \Psi_{k_j}(\|B_{k_j}\|)\leq X\right]\\
&=\mathbb{E}_X\left[\prod_{k=1}^K\mathbb{E}_{\Phi_k}\left(\prod_{B_{k_j}\in\Phi_k} \mathbb{P}\left[\Psi_k(\|B_{k_j}\|)\leq X\bigg| X, \Phi_k\right]\right)\right]\\
&\stackrel{(a)}{=}\mathbb{E}_X\left[\prod_{k=1}^K \exp\left(-2\pi \lambda_k \int^{\infty}_0 \mathbb{P}[\Psi_k(r)\geq X\big|X] r\dif r\right)\right]\\
&\stackrel{(b)}{=}\mathbb{E}_X\left[\exp\left(-\pi\sum_{k=1}^K\lambda_k\int^{\infty}_0\mathbb{P}\left[y\leq\left(\Psi_k^{-1}(X)\right)^2\big|X\right]\dif y\right)\right]\\
\end{align*}
\begin{align*}
&\stackrel{(c)}{=}\mathbb{E}_X\left[\exp\left(-\pi\sum_{k=1}^K\lambda_k\mathbb{E}_{\Psi_k}\left[\left(\Psi_k^{-1}(X)\right)^2\big| X\right]\right)\right],
\end{align*}
where $(a)$ follows from the probability generating functional (PGFL) of $K$ independent homogeneous PPPs\cite{MHRKG09,DSWKJM96}, $(b)$ is due to the fact that we let $y=r^2$ and $\Psi_k(\cdot)$ is a monotonic decreasing and invertible function, and $(c)$ is obtained since $\int^{\infty}_0\mathbb{P}\left[y\leq\left(\Psi_k^{-1}(X)\right)^2\big|X\right]\dif y=\mathbb{E}_{\Psi_k}\left[\left(\Psi_k^{-1}(X)\right)^2\big| X\right]$. Thus, the result in \eqref{Eqn:DisMaxAssFun1} is obtained. Now consider $X$ is correlated with $\Psi_{k_j}$ and thus we have
\begin{align*}
&\mathbb{E}\left[F_{\Psi^*(\|B^*\|)}(X)\right]=\mathbb{P}\left[\sup_{B_{k_j}\in\bigcup_{k=1}^K\Phi_k} \Psi_{k_j}(\|B_{k_j}\|)\leq X\right]\\
&=\prod_{k=1}^K\mathbb{E}_{\Phi_k,\Psi_k,X}\left\{\prod_{B_{k_j}\in\Phi_k} \mathbb{P}\left[\Psi_{k}(\|B_{k_j}\|)\leq X|\Psi_k,X\right]\right\}\\
&=\mathbb{E}\left\{\exp\left(-2\pi\sum_{k=1}^K \lambda_k\int_{0}^{\infty}\mathbb{P}\left[\Psi_{k}(x)\geq X|\Psi_k,X\right]x\dif x\right)\right\}\\
&=\mathbb{E}\left\{e^{-\pi\sum_{k=1}^K \lambda_k\int_{0}^{(\Psi^{-1}_k(X))^2}\mathbb{P}\left[y\leq (\Psi^{-1}_k(X))^2|\Psi_k,X\right]\dif y}\right\},
\end{align*}
which equals to \eqref{Eqn:DisMaxAssFun2} since $\mathbb{P}\left[y\leq (\Psi^{-1}_k(X))^2|\Psi_k,X\right]=1$ for $y\in[0,(\Psi^{-1}_k(X))^2]$.

\subsection{Proof of Theorem \ref{Thm:CDFDisAssBS}}\label{App:ProofCDFDisAssBS}
First consider all UCA functions are random. According to \eqref{Eqn:DisMaxAssFun1}, the probability that $B^*$ belongs to $\Phi_k$ can be written as
\begin{align*}
\vartheta_k&=\mathbb{P}\left[\sup_{B_{k_j}\in\Phi_k}\Psi_k(\|B_{k_j}\|) \geq \sup_{B_{m_j}\in\bigcup_{m\in\mathcal{K}\setminus k}\Phi_m}\Psi_m(\|B_{m_j}\|)\right]\\
&=\mathbb{E}_{Z_k}\left[\exp\left(-\pi\sum_{m\in\mathcal{K}\setminus k} \lambda_m \mathbb{E}_{\Psi_m}\left[\left(\Psi^{-1}_m\left(Z_k\right)\right)^2\big|Z_k\right]\right)\right],
\end{align*}
where $Z_k=\sup_{B_{k_j}\in\Phi_k}\Psi_k(\|B_{k_j}\|)\defn \Psi_k(\|B^*_k\|)$ for $B^*_k\defn\arg\sup_{B_{k_j}\in\Phi_k}\Psi_k(\|B_{k_j}\|)$. Applying Theorem \ref{Thm:CDFMaxAssFun} to the network with one tier, the CDF of $Z_m$ can be found as $F_{Z_k}(z)=e^{-\pi\lambda_k\mathbb{E}_{\Psi_k}\left[\left(\Psi_k^{-1}(z)\right)^2\right]}$ and thus the pdf of $Z_k$ is given by
\begin{align*}
f_{Z_k}(z)=& 2\pi\lambda_k\mathbb{E}_{\Psi_k}\left[-\Psi^{-1}_k(z)\frac{\dif \Psi^{-1}_k(z)}{\dif z}\right]\times\\ &\exp\left(-\pi\lambda_k\mathbb{E}_{\Psi_k}\left[\left(\Psi_k^{-1}(z)\right)^2\right]\right)
\end{align*}
and note that $\frac{\dif\Psi^{-1}_k(z)}{\dif z}<0$. Therefore, it follows that
\begin{align}
\vartheta_k =&\bigintsss_0^{\infty}\exp\left(-\pi\sum_{k\in\mathcal{K}\setminus m} \lambda_k \mathbb{E}_{\Psi_k}\left[\left(\Psi^{-1}_k\left(z\right)\right)^2\right]\right)f_{Z_k}(z)\dif z\nonumber\\
=& -2\pi\lambda_k\mathbb{E}_{\Psi_k}\bigg[\int_0^{\infty}e^{-\pi\sum_{m\in\mathcal{K}} \lambda_m\mathbb{E}_{\Psi_m}\left[\left(\Psi^{-1}_m\left(z\right)\right)^2\right]}\times\nonumber\\
&\hspace{0.8in}\Psi^{-1}_m(z)\dif \Psi^{-1}_k(z)\bigg], \label{Eqn:AssProbTierMProof}
\end{align}
which equals to \eqref{Eqn:CellAssociationProb1}.
Now if all the UCA functions are deterministic, then using the total probability law the CDF of $\|B^*_0\|$ can be written as
\begin{align}
F_{\|B^*_0\|}(x)&=\sum_{m\in\mathcal{K}} \mathbb{P}\left[\|B^*_0\|\leq x | B^*_0\in\Phi_m\right]\mathbb{P}[B^*_0\in\Phi_m]\nonumber\\ &\stackrel{(a)}{=}\sum_{m\in\mathcal{K}} \mathbb{P}\left[\Psi_m(\|B^*_0\|)\geq \Psi_m(x) | B^*_0\in\Phi_m\right]\vartheta_k\nonumber\\
&\stackrel{(b)}{=}1-\sum_{k\in\mathcal{K}}\exp\left(-\pi\sum_{m\in\mathcal{K}}\lambda_m\left[\Psi_m^{-1}\circ\Psi_k(x)\right]^2\right)\vartheta_k, \label{Eqn:CDFAssBSProof}
\end{align}
where $(a)$ is due to the fact that $\Psi_k(\cdot)$ is monotonic decreasing and $(b)$ follows from the result in \eqref{Eqn:DisMaxAssFun2}.  Substituting \eqref{Eqn:AssProbTierMProof} into  \eqref{Eqn:CDFAssBSProof} results in \eqref{Eqn:CDFDistAssBS}.

\subsection{Proof of Lemma \ref{Lem:CellLoadStat}}\label{App:ProofCellLoadStat}
According to Theorem \ref{Thm:CDFMaxAssFun}, the CCDF of the maximum association function $\Psi^*(\|B^*\|)$ is 
\begin{align*}
\mathbb{P}\left[\Psi^*(\|B^*\|)\geq \frac{1}{x^{\alpha}}\right]&=1-\mathbb{P}\left[\sup_{B_{k_j}\in\bigcup_{k\in\mathcal{K}}\Phi_k}\frac{\psi_{k_j}}{\|B_{k_j}\|^{\alpha}}\leq \frac{1}{x^{\alpha}}\right]\\
&=1-\exp\left(-\pi x^2\sum_{k=1}^{K}\lambda_k\mathbb{E}\left[\psi^{\frac{2}{\alpha}}_k\right]\right),
\end{align*}
which indicates $\Psi^*(B^*)$ is statistically equal to be the distance from the BS in the
$K$ independent homogeneous PPPs nearest to the typical user, i.e., $\Psi^*(B^*)\stackrel{d}{=}\inf_{\tilde{B}_{k_j}\in\bigcup_{k\in\mathcal{K}}\tilde{\Phi}_k}\|\tilde{B}_{k_j}\|$ where $\stackrel{d}{=}$ denotes the equivalence in distribution and  $\tilde{\Phi}_k\defn\{\tilde{B}_{k_j}\in\mathbb{R}^2: \tilde{B}_{k_j}=\psi^{-\frac{1}{\alpha}}_{k_j}B_{k_j},B_{k_j}\in\Phi_k, j\in\mathbb{N}_+\}$ is the $k$th homogeneous PPP of intensity $\lambda_k\mathbb{E}\left[\psi^{\frac{2}{\alpha}}_k\right]$. Let $\tilde{B}^*_k$ denote the point in $\tilde{\Phi}_k$ nearest to the typical user and the probability that $B^*$ is from the $k$th tier can be found as
\begin{align*}
\vartheta_k&=\mathbb{P}\left[\tilde{B}_k^*\in\tilde{\Phi}_k\right]=\mathbb{P}\left[\|\tilde{B}^*_k\|\leq \min_{m\in\mathcal{K}\setminus k}\{\|\tilde{B}^*_m\|\}\right]\\
&=\prod_{m\in\mathcal{K}\setminus k}\mathbb{P}\left[\|\tilde{B}^*_k\|\leq \|\tilde{B}^*_m\|\right]\\
&=\mathbb{E}\left[\exp\left(-\pi\|\tilde{B}^*_k\|^2\sum_{m\in\mathcal{K}\setminus k}\lambda_m\mathbb{E}\left[\psi^{\frac{2}{\alpha}}_m\right]\right)\right]\\
&=\frac{\lambda_k\mathbb{E}\left[\psi^{\frac{2}{\alpha}}_k\right]}{\sum_{m=1}^{K}\lambda_m\mathbb{E}\left[\psi^{\frac{2}{\alpha}}_m\right]}.
\end{align*}
Since each user independently decides with which BS it associates, the users that associate with a tier-$k$ BS, called tier-$k$ users, is a thinning homogeneous PPP of intensity $\lambda_{\subU}\mathbb{P}\left[B^*\in\Phi_k\right]=\lambda_{\subU}\vartheta_k$.

Now we can image that the tier-$k$ users are distributed over the entire plane consisting of the Voronoi-tessellated cells of the BSs in $\tilde{\Phi}_k$ in that the users associate with their nearest BS in $\tilde{\Phi}_k$. In other words, $p_{k,n}$ in \eqref{Eqn:DefnNumUserTierkBS} can be equivalently expressed as
\begin{align}
p_{k,n}=\frac{1}{n!}\mathbb{E}\left[\left(\frac{\lambda_{\subU}\lambda_k}{\widetilde{\lambda}_k}\nu(\widetilde{C}_k)\right)^n e^{-\frac{\lambda_{\subU}\lambda_k}{\widetilde{\lambda}_k}\nu(\widetilde{C}_k)}\right],
\end{align}
where $\widetilde{C}_k$ is the Voronoi cell of a BS in $\tilde{\Phi}_k$. Although the exact distribution of the Lebesgue measure of a Voronoi cell is still an open problem, it can be accurately approximated by a gamma distribution with appropriate parameters\cite{JSFNZ07}. According to References \cite{JSFNZ07,CHLLCW16}, we learn that the accurate pdf of $\nu(\widetilde{C}_k)$ can be inferred as
\begin{align}
f_{\nu(\widetilde{C}_k)} (x) = \frac{\left(\zeta_k\lambda_k x\right)^{\zeta_k}}{x\Gamma(\zeta_k)}e^{-\zeta_kx\lambda_k}.
\end{align}   
Thus, it follows that
\begin{align*}
p_{k,n}= \frac{\left(\zeta_k\lambda_k\right)^{\zeta_k}}{n!\Gamma(\zeta_k)}\left(\frac{\lambda_{\subU}\lambda_k}{\widetilde{\lambda}_k}\right)^n \int_{0}^{\infty} x^{\zeta_k+n-1} e^{-\left(\frac{\lambda_{\subU}}{\widetilde{\lambda}_k}+\zeta_k\right)\lambda_k x} \dif x,
\end{align*}
which exactly equals to \eqref{Eqn:ProbNumUserTierkBS}. The tier-$k$ cell load can be found as
\begin{align*}
\mathbb{E}\left[\Phi_{\subU}(\mathcal{C}_k)\right]&=\sum_{n=1}^{\infty} np_{k,n}= \int_{0}^{\infty}\left(\lambda_{\subU}\mathbb{P}\left[B^*\in\Phi_k\right]\right) x f_{\nu(\tilde{C}_k)} (x) \dif x\\
&=\frac{\lambda_{\subU}\lambda_k\mathbb{E}\left[\psi^{\frac{2}{\alpha}}_k\right]}{\sum_{m=1}^{K}\lambda_m\mathbb{E}\left[\psi^{\frac{2}{\alpha}}_m\right]}\times \frac{1}{\lambda_k},
\end{align*}
which is the result in \eqref{Eqn:CellLoadTierk} and the proof is complete. 

\subsection{Proof of Lemma \ref{Lem:EqnGCA}}\label{App:ProofEqnGCA}
According to the energy efficiency function defined in \eqref{Eqn:DefnGCAFunction}, the GCA scheme in \eqref{Eqn:DefnGCA} can be equivalently expressed as follows
\begin{align*}
B^*&=\argsup_{B_{k_j}\in\bigcup_{k=1}^K \Phi_k} \left(1+\frac{P_kH_{k_j}}{I_{k_j}\|B_{k_j}\|^{\alpha}}\right)^{\frac{1}{\Upsilon^{\on}_k}}\\
&\stackrel{(a)}{=}\argsup_{B_{k_j}\in\bigcup_{k=1}^K \Phi_k}\bigg[1+\left(\frac{P_kH_{k_j}/\Upsilon^{\on}_k}{I_{k_j}\|B_{k_j}\|^{\alpha}}\right)\\
&\hspace{1.15in}+\mathcal{O}\left(\frac{P_kH_{k_j}/\Upsilon^{\on}_k}{I_{k_j}\|B_{k_j}\|^{\alpha}}\right)\bigg],
\end{align*}
where $(a)$ follows from the binomial expansion of $(1+X)^r=\sum_{j=0}^{\infty}{r\choose j}X^j$, $\mathcal{O}(X)$ denotes the linear sum of the higher order terms of $X$. Since all terms in $\mathcal{O}(\cdot)$ consists of the first term $\frac{P_kH_{k_j}/\Upsilon^{\on}_k}{I_{k_j}\|B_{k_j}\|^{\alpha}}$, removing them and constant $1$ does not affect the result of finding $B^*$ since $\|B^*\|$ that maximizes the first term $\frac{P_kH_{k_j}/\Upsilon^{\on}_k}{I_{k_j}\|B_{k_j}\|^{\alpha}}$ also maximizes the higher order terms in $\mathcal{O}(\cdot)$. Hence, we have
\begin{align*}
B^*&=\argsup_{B_{k_j}\in\bigcup_{k=1}^K \Phi_k}\left(\frac{P_kH_{k_j}}{\Upsilon^{\on}_kI_{k_j}\|B_{k_j}\|^{\alpha}}\right).
\end{align*}  
In addition, note that $I_{k_j}$ is the interference at the typical user for a given point set $\bigcup_{k=1}^K \Phi_k$ and $B_{k_j}$. All $I_{k_j}$'s are identical in that they are completely correlated and has the same distribution based on the Slivnyak theorem. Therefore, removing $I_{k_j}$ from $\left(\frac{P_kH_{k_j}}{\Upsilon^{\on}_kI_{k_j}\|B_{k_j}\|^{\alpha}}\right)$ does not affect the result of finding $B^*$ as well, which yields the result in \eqref{Eqn:GCA}.

\subsection{Proof of Theorem \ref{Thm:GreenCoverageProbGCA}}\label{App:ProofGreenCoverageProbGCA}
Consider users use the GCA in scheme in \eqref{Eqn:GCA} to associate with their serving BSs. According to Theorem \ref{Thm:CDFMaxAssFun} and $\Psi^{-1}_{k_j}(x)=\left(P_kH_{k_j}/\Upsilon^{\on}_k\right)^{\frac{1}{\alpha}}x^{-\frac{1}{\alpha}}$, we have the following:
\begin{align*}
&1-F_{\Psi^*(\|B^*\|)}(x)=\mathbb{P}\left[\Psi^*(\|B^*\|)\geq x\right]\\
&=1-\mathbb{P}\left[\sup_{B_{k_j}\in\bigcup_{k=1}^K\Phi_k} \frac{P_kH_{k_j}}{\Upsilon^{\on}_k\|B_{k_j}\|^{\alpha}}\leq x\right]\\
&=1-\exp\left(-\pi x^{-\frac{2}{\alpha}}\sum_{k=1}^{K}\lambda_k\left(\frac{P_k}{\Upsilon^{\on}_k}\right)^{\frac{2}{\alpha}}\mathbb{E}\left[H_k^{\frac{2}{\alpha}}\right]\right)\\
&=1-\exp\left(-\pi y^2\sum_{k=1}^{K}\widehat{\lambda}_k \right)=\mathbb{P}\left[\|\widehat{B}^*\|\leq y\right],
\end{align*}
where $\|\widehat{B}^*\|\defn(\Psi(\|B^*\|))^{-\frac{1}{\alpha}}$, $\widehat{\lambda}_k=\lambda_k\left(\frac{P_k}{\Upsilon^{\on}_k}\right)^{\frac{2}{\alpha}}\mathbb{E}\left[H_k^{\frac{2}{\alpha}}\right]$ and $y\defn x^{-\frac{1}{\alpha}}$. Thus, $\widehat{B}^*\in \bigcup_{k=1}^K\widehat{\Phi}_k$ can be viewed as the point in $\bigcup_{k=1}^K\widehat{\Phi}_k$ nearest to the typical user where $\widehat{\Phi}_k$ is a homogeneous PPP of intensity $\widehat{\lambda}_k$ and all $\widehat{\Phi}_k$'s are independent. Moreover, for $B^*\in\Phi_k$ we can have the following
\begin{align*}
\frac{H_kP_k}{I^*\|B^*\|^{\alpha}}&=\frac{\Upsilon^{\on}_k\Psi^*(\|B^*\|)}{\sum_{B_{m_j}\in\bigcup^K_{m=1}\Phi_k\setminus B^*}\Upsilon^{\on}_mV_{m_j}\Psi_{m_j}(\|B_{m_j}\|)}\\
&\stackrel{d}{=}\frac{\Upsilon^{\on}_k}{\widehat{I}^* \|\widehat{B}^*\|^{\alpha}},
\end{align*}
where $\widehat{I}^*\defn \sum_{\widehat{B}_{k_j}\in\bigcup^K_{m=1}\widehat{\Phi}_m\setminus\widehat{B}^*}\Upsilon^{\on}_mV_{m_j}\|\widehat{B}_{m_j}\|^{-\alpha}$ and $\stackrel{d}{=}$ means the equivalence in distribution. Hence, it follows that 
\begin{align}
\rho_{\gc}(\eta)&=\sum_{k=1}^{K} \mathbb{P}\left[\frac{P_kH_k}{I^*\|B^*\|^{\alpha}}\geq 2^{\eta \Upsilon^{\on}_k}-1\bigg|B^*\in\Phi_k\right]\vartheta_k\nonumber\\
&=\sum_{k=1}^{K}\mathbb{P}\left[\frac{\Upsilon^{\on}_k}{\widehat{I}^* \|\widehat{B}^*\|^{\alpha}}\geq 2^{\eta \Upsilon^{\on}_k}-1\right]\vartheta_k\nonumber\\
&=\sum_{k=1}^{K} \mathbb{P}\left[\frac{(2^{\eta \Upsilon^{\on}_k}-1)\|\widehat{B}^*\|^{\alpha}\widehat{I}^*}{\Upsilon^{\on}_k}\leq 1\right]\vartheta_k\nonumber\\
&\stackrel{(a)}{=}\sum_{k=1}^{K}\vartheta_k\mathcal{L}^{-1}\left\{\frac{1}{s}\mathbb{E}\left[e^{-s\frac{(2^{\eta \Upsilon^{\on}_k}-1)}{\Upsilon^{\on}_k}\|\widehat{B}^*\|^{\alpha}\widehat{I}^*}\right]\right\}\left(1\right), \label{Eqn:GreenCovProbProof}
\end{align}
where $(a)$ follows from the identity that $\mathbb{P}[Z\leq z]=\mathcal{L}^{-1}\left\{s^{-1}\mathbb{E}[e^{-sZ}]\right\}(z)$. Using the algebraic technique in the proof of Proposition 2 in \cite{CHLLCW16} and letting $s_{m_k}=s \frac{\Upsilon^{\on}_m(2^{\eta \Upsilon^{\on}_k}-1)}{\Upsilon^{\on}_k}$, the Laplace transform of $\|\widehat{B}^*\|^{\alpha}\widehat{I}^*$ can be found as follows
\begin{align*}
&\mathbb{E}\left[\exp\left\{-s\frac{\left(2^{\eta \Upsilon^{\on}_k}-1\right)}{\Upsilon^{\on}_k}\|\widehat{B}^*\|^{\alpha}\widehat{I}^*\right\}\right]\\
&\stackrel{(b)}{\gtrapprox} \prod_{m=1}^{K}\mathbb{E}\left[e^{-\sum_{\widehat{B}_{m_j}\in\widehat{\Phi}_k\setminus\widehat{B}^*}s_{m_k}V_{m_j} \left(\frac{\|\widehat{B}^*\|}{\|\widehat{B}_{m_j}\|}\right)^{\alpha}}\right]\\
&\stackrel{(c)}{=} \mathbb{E}\bigg[\exp\bigg(-\pi \|\widehat{B}^*\|^2\sum_{m=1}^{K}\widehat{\lambda}_m(1-p_{m,\subVoid})\times\\
&\hspace{0.25in}\int_{1}^{\infty}\left(1-e^{-s_{m_k} x^{-\frac{\alpha}{2}}}\right)\dif x\bigg)\bigg],
\end{align*}
where $(b)$ follows by assuming all $V_{m_j}$'s are independent and the non-void BS are independent PPPs that give rise to slightly larger interference so that this lower bound is fairly tight \cite{CHLLCW15,CHLLCW16}, $(c)$ follows from the probability generating functional of $K$ independent homogeneous PPPs\cite{MHRKG09,DSWKJM96}, and the integral can be simplified as follows
\begin{align*}
&\int_{1}^{\infty}\left(1-e^{-s_{m_k} x^{-\frac{\alpha}{2}}}\right)\dif x=\int_{0}^{\infty}\left(1-e^{-s_{m_k} x^{-\frac{\alpha}{2}}}\right)\dif x\\
&-\int_{0}^{1}\left(1-e^{-s_{m_k} x^{-\frac{\alpha}{2}}}\right)\dif x\stackrel{(d)}{=} \int_{0}^{\infty} \mathbb{P}\left[x\leq \left(\frac{s_{m_k}}{Y}\right)^{\frac{2}{\alpha}}\right] \dif x-1\\
&+\frac{2}{\alpha}s^{\frac{2}{\alpha}}_{m_k}\int_{s_{m_k}}^{\infty} u^{-(1+\frac{2}{\alpha})} e^{-u} \dif u=s_{m_k}^{\frac{2}{\alpha}}\bigg[\Gamma\left(1-\frac{2}{\alpha}\right)+\\
&\frac{2}{\alpha}\Gamma\left(-\frac{2}{\alpha},s_{m_k}\right)\bigg]-1=\hbar\left(s_{m_k},\frac{2}{\alpha}\right),
\end{align*}
where $(d)$ follows from assuming $Y$ is an exponential random variable with unit mean and $\Gamma(z,x)=\int_{x}^{\infty}t^{z-1}e^{-t}\dif t$ is the lower incomplete gamma function. Then we can have
\begin{align}
\mathbb{E}\left[e^{-s\frac{(2^{\eta \Upsilon^{\on}_k}-1)}{\Upsilon^{\on}_k}\|\widehat{B}^*\|^{\alpha}\widehat{I}^*}\right]\gtrapprox \bigg(1+\sum_{m=1}^{K}(1-p_{m,\subVoid})\times\nonumber\\
\hbar\left(s_{m_k},\frac{2}{\alpha}\right)\vartheta_m\bigg)^{-1} \label{Eqn:LaplaceTransInterSigRatio}
\end{align}
since  $f_{\|\widehat{B}^*\|}(x)=2\pi(\sum_{k=1}^{K}\widehat{\lambda}_k)xe^{-\pi x^2\sum_{k=1}^{K}\widehat{\lambda}_k}$ and substituting \eqref{Eqn:LaplaceTransInterSigRatio} into \eqref{Eqn:GreenCovProbProof} yields the following result
\begin{align}
\rho_{\gc}(\eta)\gtrapprox& \sum_{k=1}^{K}\vartheta_k\mathcal{L}^{-1}\bigg\{s^{-1}\bigg(1+\sum_{m=1}^{K}(1-p_{m,\subVoid})\times\nonumber\\
&\hbar\left(s\frac{\Upsilon^{\on}_m(2^{\eta\Upsilon^{\on}_k}-1)}{\Upsilon^{\on}_k},\frac{2}{\alpha}\right)\vartheta_m\bigg)^{-1}\bigg\}\left(1\right),
\end{align}
which equals to \eqref{Eqn:GreenCoverProbGCA} due to $ \int_{0}^{t}g(\tau)\dif\tau=\mathcal{L}^{-1}\{G(s)/s\}(t)$.

\subsection{Proof of Theorem \ref{Thm:GreenCoveProbGCA2}}\label{App:ProofGreenCoveProbGCA2}
According to the proof of Theorem \ref{Thm:GreenCoverageProbGCA}, we also can know the SIR in \eqref{Eqn:DefnGreenCoverage} has the following equivalence in distribution:
\begin{align*}
\frac{H^f_kQ_kP_k}{I^*\|B^*\|^{\alpha}}&=\frac{H^f_k\Upsilon^{\on}_k\Psi^*(\|B^*\|)}{\sum_{B_{m_j}\in\bigcup^K_{m=1}\Phi_k\setminus B^*}H^f_{m_j}\Upsilon^{\on}_mV_{m_j}\Psi_{m_j}(\|B_{m_j}\|)}\\
&\stackrel{d}{=}\frac{H^f_k\Upsilon^{\on}_k}{\widehat{I}^* \|\widehat{B}^*\|^{\alpha}},
\end{align*}
where $\widehat{I}^*\defn \sum_{\widehat{B}_{m_j}\in\bigcup^K_{m=1}\widehat{\Phi}_m\setminus\widehat{B}^*}H^f_{m_j}\Upsilon^{\on}_mV_{m_j}\|\widehat{B}_{m_j}\|^{-\alpha}$ and $\widehat{\Phi}_k$ is a PPP of intensity $\widehat{\lambda}_k=(P_k/\Upsilon^{\on}_k)^{\frac{2}{\alpha}}\mathbb{E}\left[Q_k^{\frac{2}{\alpha}}\right]\lambda_k$. Thus, it follows that
\begin{align*}
\rho_{\gc}(\eta)&=\sum_{k=1}^{K}\mathbb{P}\left[\frac{H^f_k\Upsilon^{\on}_k}{\widehat{I}^* \|\widehat{B}^*\|^{\alpha}}\geq 2^{\eta \Upsilon^{\on}_k}-1\right]\vartheta_k\\
&=\sum_{k=1}^{K} \mathbb{E}\left [\exp\left(-\frac{(2^{\eta\Upsilon^{\on}_k}-1)}{\Upsilon^{\on}_k}\widehat{I}^* \|\widehat{B}^*\|^{\alpha}\right)\right]\vartheta_k.
\end{align*}
Now consider all $\widehat{\Phi}_k$'s are $K$ independent PPPs and the Laplace transform of $\widehat{I}^*\|\widehat{B}^*\|^{\alpha}$ with parameter $(2^{\eta \Upsilon^{\on}_k})/\Upsilon^{\on}_k$ can be found by using the proof of Proposition 2 in \cite{CHLLCW16}. Accordingly, we have
\begin{align*}
\rho_{\gc}(\eta) \gtrapprox& \sum_{k=1}^{K}\vartheta_k\mathbb{E}\bigg[ \exp\bigg(-\pi\|\widehat{B}^*\|^2\times\\
&\sum_{m=1}^{K}\ell\left(\frac{\Upsilon^{\on}_m(2^{\eta\Upsilon^{\on}_k}-1)}{\Upsilon^{\on}_k},\frac{2}{\alpha}\right)(1-p_{m,\subVoid})\widehat{\lambda}_k\bigg) \bigg] 
\end{align*} 
and then averaging this tight lower bound over $\|\widehat{B}^*\|^2$ with $f_{\|\widehat{B}^*\|}(x)=2\pi(\sum_{k=1}^{K}\widehat{\lambda}_k)e^{-\pi x^2 (\sum_{k=1}^{K}\widehat{\lambda}_k)}$ results in \eqref{Eqn:GreenCoveProbGCA2}.

\subsection{Proof of Theorem \ref{Thm:GCAEnergyEff}}\label{App:ProofGCAEnergyEff}
Although $\delta_{\gc}$ can be found by integrating $\rho_{\gc}(\eta)$ in \eqref{Eqn:GreenCoverProbGCA} with respect to $\eta$ over $[0,\infty]$, this method is unable to yield a much neat result. Here we find $\delta_{\gc}$ by calculating the mean spectrum efficiency of the associated BS in a more tractable way shown in the following
\begin{align*}
\mathds{C}_{\gc,k}&=\frac{1}{\ln 2}\mathbb{E}\left[\ln\left(1+\frac{P_kH_k}{I^*\|B^*\|^{\alpha}}\right)\right]\nonumber\\
&=\frac{1}{\ln 2}\int_{0^+}^{1} \mathbb{E}\left[\left(\frac{I^*\|B^*\|^{\alpha}}{P_kH_k}+y\right)^{-1}\right]\dif y.
\end{align*}
Also, we know
\begin{align*}
&\mathbb{E}\left[\left(\frac{I^*\|B^*\|^{\alpha}}{P_kH_k}+y\right)^{-1}\right]\stackrel{(a)}{=}\mathbb{E}\left[\left(\frac{\widehat{I}^*\|\widehat{B}^*\|^{\alpha}}{\Upsilon^{\on}_k}+y\right)^{-1}\right]\\
&=\int_{0^+}^{\infty} e^{-sy}\mathbb{E}\left[e^{-s\frac{\widehat{I}^*\|\widehat{B}^*\|^{\alpha}}{\Upsilon^{\on}_k}}\right]\dif s\\
&\stackrel{(b)}{\gtrapprox} \bigintsss_{0^+}^{\infty} e^{-sy} \left(1+\sum_{m=1}^{K}(1-p_{m,\subVoid})\hbar\left(\frac{s\Upsilon^{\on}_m}{\Upsilon^{\on}_k},\frac{2}{\alpha}\right) \vartheta_m\right)^{-1}\dif s
\end{align*}
where $(a)$ follows from the definitions of $\widehat{I}^*$ and $\widehat{B}^*$ in the proof of Theorem \ref{Thm:GreenCoverageProbGCA} and $(b)$ is obtained by applying the result in \eqref{Eqn:LaplaceTransInterSigRatio}. Therefore, it follows that
\begin{align*}
&\mathbb{E}\left[\left(\frac{I^*\|B^*\|^{\alpha}}{P_kH_k}+y\right)^{-1}\right]\gtrapprox \frac{1}{\ln 2} \int_{0^+}^{\infty} \int_{0}^{1}e^{-sy}\dif y\times\\ &\left(1+\sum_{m=1}^{K}(1-p_{m,\subVoid})\hbar\left(\frac{s\Upsilon^{\on}_m}{\Upsilon^{\on}_k},\frac{2}{\alpha}\right) \vartheta_m\right)^{-1}\dif s\\
&= \frac{1}{\ln 2} \bigintsss_{0^+}^{\infty} \frac{(1-e^{-s})\dif s}{s\left(1+\sum_{m=1}^{K}(1-p_{m,\subVoid})\hbar\left(\frac{s\Upsilon^{\on}_m}{\Upsilon^{\on}_k},\frac{2}{\alpha}\right) \vartheta_m\right)}
\end{align*}
and then substituting this result into \eqref{Eqn:DefnLinkEngEffic} yields \eqref{Eqn:GCALinkEnergyEff}.

\bibliographystyle{ieeetran}
\bibliography{IEEEabrv,Ref_GreenUserAss}


\end{document}